\documentclass[trackchanges, twocolumn]{aastex701}
\usepackage{amsmath}
\usepackage{bm}

\setcitestyle{comma}
\begin{document}

\title{The impact of baryons on weak lensing statistics as a function of halo mass and radius}

\author[0000-0002-2318-3087]{Max E. Lee}
\affiliation{Department of Astronomy, Columbia University, MC 5246, 538 West 120th Street, New York, NY 10027, USA}
\email[show]{max.e.lee@columbia.edu}  

\author[0000-0003-3633-5403]{Zolt\'an Haiman}
\affiliation{Department of Astronomy, Columbia University, MC 5246, 538 West 120th Street, New York, NY 10027, USA}
\affiliation{Department of Physics, Columbia University, MC 5255, 538 West 120th Street, New York, NY 10027, USA}
\affiliation{Institute of Science and Technology Austria, Am Campus 1, Klosterneuburg 3400 Austria}
\email{Zoltan.Haiman@ista.ac.at}

\author[0000-0002-3185-1540]{Shy Genel}
\affiliation{Center for Computational Astrophysics, Flatiron Institute, 162 Fifth Ave, New York, NY, 10010, USA}
\affiliation{Columbia Astrophysics Laboratory, Columbia University, 550 West 120th Street, New York, NY 10027, USA}
\email{sgenel@flatironinstitute.org}

\begin{abstract}
\noindent Upcoming weak lensing (WL) surveys such as those by {\it Euclid}, LSST, and {\it Roman} require percent-level control over systematic effects. A common approach to mitigating baryonic effects uses semi-analytic baryon correction models (BCMs) that modify halo profiles in dark matter-only (DMO) simulations, calibrated to statistics from hydrodynamic simulations. We investigate the limits of this approach by progressively replacing larger regions around halos of decreasing mass in DMO simulations with their hydrodynamical counterparts. We compare multiple statistics --- the matter ($P(k)$) and weak-lensing ($C_\ell$) power spectra, peak counts, minima, one-point PDFs, and Minkowski functionals --- from ``Replace" fields against hydrodynamical and DMO simulations. We find that replacing all halos with $M\geq10^{12}\,h^{-1}\,{\rm M}_\odot$ out to $r\leq5R_{200}$ recovers $\sim 90\%$ of the baryonic suppression in $P(k)$ and $C_\ell$ with the remaining $\sim 10\%$ originating from lower-mass halos or material farther outside of DM halos. Each statistic has distinct sensitivities to baryons: $P(k)$ and $C_\ell$ are sensitive to a broad range of masses and radii, whereas WL peaks are primarily affected by the cores of massive halos. We show that BCMs applied to massive halos and calibrated to match hydrodynamical $P(k)$ make two cancelling ``mistakes": they underpredict core masses and compensate by overpredicting baryonic impacts at larger radii, thereby explaining previously reported failures of peak statistics in these models. We provide a framework for diagnosing critical mass/radius regions in baryonic modeling for a range of statistics for next-generation BCMs.
\end{abstract}

% \keywords{\uat{Galaxies}{573}}


\section{Introduction}
\label{sec:intro}

Realizing the full scientific potential of upcoming weak gravitational lensing surveys such as \textit{Euclid} \citep{Euclid11}, the Vera C. Rubin Observatory’s Legacy Survey of Space and Time (LSST) \citep{LSST}, and the Nancy Grace \textit{Roman} Space Telescope \citep{Roman15} requires percent-level control of systematics. This is complicated by baryonic feedback, which redistributes matter within and around dark matter halos \citep{schaye10, McCarthy17, Volker10, Nelson2019, hadzhiyska_evidence_2024, bigwood2024weaklensingcombinedkinetic} suppressing the matter power spectrum by 10–20\% at scales $k \sim 10\,h\,\mathrm{Mpc}^{-1}$ \citep{Arico2020, Osato21, Robertson2026} and altering non-Gaussian statistics, such as weak lensing peak counts, minima, and Minkowski functionals \citep{jain00, Liu15, Chisari19, Martinet2021, leeComparingWeakLensing2022}, which contain a wealth of information and constraining power \citep{Jeffrey2025}. If left untreated, these effects can bias cosmological parameter constraints by amounts exceeding the statistical uncertainties of Stage-IV data \citep{Hikage19, Hamana20, Amon-2020, DES21, Abbott22, Abbot-26}.

Modelling baryonic effects and their uncertainties remains a challenge. Hydrodynamical simulations self-consistently include gas physics, star formation, and feedback via parameterized models typically referred to as subgrid physics, which have been shown to accurately reproduce various observables sensitive to baryonic effects, such as stellar distributions \citep{Pillepich17}. However, hydrodynamical simulations remain computationally prohibitive for the large volumes required by modern surveys \citep{schaye10, Pillepich17, Nelson2019}. A single flagship hydrodynamical simulation run, such as IllustrisTNG or FLAMINGO, requires tens of millions of CPU hours \citep{Nelson2019, Flamingo-23} and typically provides a single simulation realization at one or a few locations in the high-dimensional subgrid model parameter space, making marginalization over baryonic effects difficult\footnote{Subgrid models such as the ones from IllustrisTNG can have upwards of 28 parameters \citep{Ni2023}. The CAMELS suites of simulations have performed systematic explorations of this space along with other subgrid parameter models in \citep{camels20, Ni2023}, but are limited to smaller cosmological box sizes.}. 

This computational expense has motivated the development of semi-analytic empirical \emph{baryonic correction models} (BCMs), which attempt to capture baryonic effects by modifying dark-matter-only (DMO) simulation particles with prescriptions for bound gas, gas ejection, stellar fractions, and dark matter relaxation \citep{Schneider-15, Schneider19, Arico2020, Arico20b, Anbajagane-24, Schneider-25}. BCM parameters can be calibrated to match statistics measured in hydrodynamical simulations \citep{Arico+2021,Zhou25}, or fit to observables such as Sunyaev–Zel’dovich (SZ) and X-ray data \citep{Schneider19, Schneider-25, Grandis-24}, making them powerful and physically motivated alternatives to full hydrodynamical simulations.

Despite their computational efficiency, BCMs face a fundamental question of physical self-consistency: if a model is tuned to reproduce a statistic—for example, the 3D matter power spectrum $P(k)$—does it automatically reproduce the correct baryonic impact on other observables, such as the weak lensing convergence power spectrum $C_\ell$ \citep{heymans13, kitching14, hildebrandt17, Hikage19}, peak counts and minima \citep{jain00, yang11, Liu15, kacprzak16, Li2019, Coulton2020, Martinet2021, Zurcher2021, Harnois2021}, one-point convergence probability distribution functions \citep{thiele_accurate_2020}, or Minkowski functinoals \citep{Maturi+2010,Munshi11, Kratochvil+2012, petri13, Marques2019, Gatti21}?

Recent comparisons of BCMs with hydrodynamical simulations have revealed discrepancies. BCMs tuned to match $P(k)$ at the few-percent level fail to reproduce peak count distributions \citep{leeComparingWeakLensing2022} or bispectra \citep{Arico+2021} at the same level of precision, and at detectable levels for stage IV surveys. More fundamentally, \citet{Miller-25} demonstrated that an idealized baryon correction model, which directly replaces halo particles in an N-body dark matter only simulation with particles from their matched hydrodynamical counterparts out to $2R_{200}$ in IllustrisTNG, does not fully recover the hydrodynamical matter power spectrum. This discrepancy implies that some of the baryonic effect is sourced from lower-mass halos or diffuse intergalactic components beyond the mass and radius thresholds explored \citep{Miller-25, Ayromlou-23}. Moreover, if even field-level replacements in dark matter only simulations set to match the effects of hydrodynamics perfectly fail to reconstruct $P(k)$, then BCMs tuned to $P(k)$ must be compensating by placing baryonic modifications in regions of halo mass and radius space that are inconsistent with their hydrodynamical halo counterparts. If this is the case, these modifications may propagate inconsistently to other statistics.

This raises the question that this work seeks to answer: Which halo masses and halocentric radii dominate the baryonic impact on a given statistic? Understanding this is essential for determining where in mass-radius space baryonic models should provide the most accurate predictions and where approximations are more tolerable. Furthermore, answering this question provides a method to test whether a BCM is self-consistent across multiple statistics. Addressing this requires a framework to quantify how arbitrary statistics respond to baryonic modifications across halo-mass bins and radial shells.

We introduce a self-consistent formalism that addresses this by defining a family of hybrid ``Replace" fields where, for halos in the mass range $[M_a, M_b]$, N-body particles are replaced by hydrodynamical simulation particles within halocentric radial shells $[r_i, r_j]$. By applying this replacement to the IllustrisTNG simulations over a set of various mass bins and radial shell combinations, we build a map of where baryonic effects on different statistics, such as matter power spectra, weak-lensing two-point functions, peak counts, minima, PDFs, and Minkowski functionals, originate in this mass-radius space. We then consider the placement of baryons by BCMs and their influence on the above measured statistics.

This paper is organized as follows. In \S~\ref{sec:formalism}, we develop the response framework. In \S~\ref{sec:simulations}, we describe the IllustrisTNG DMO and hydrodynamical runs, the halo-matching procedure, the raytracing algorithm, and the construction of the \textit{Replace} density fields. \S~\ref{sec:results_matter} presents responses for matter power spectra, and quantifies the contribution of different mass bins and radial shells. \S~\ref{sec:results_WL} extends the analysis to weak lensing observables. We then discuss the implications for several popular baryon-correction models in the literature in \S~\ref{sec:BCM}. We summarize our findings and discuss future extensions in \S~\ref{sec:conclusion}.

\section{A Response Formalism}
\label{sec:formalism}
We begin with a general treatment of a statistic’s response to baryonic injections. We take as our inspiration the ``separate universe simulations" of \citet{Li14} and \citet{Wagner15}, in which the authors studied the response of the power spectrum in the non-linear regime to the addition of a large-scale background mode. Unlike the preceding work, we are interested in statistical responses to baryonic modifications within specific halo-mass and radius regions. To this end, we adopt a simplified, empirical approach: we treat the response of statistics to baryons as a function of mass and radius by modifying the field level of an N-body simulation to match that of a hydrodynamical simulation for the chosen set of halo mass and radii.

In general, we define a \textit{Replace} density field based on the discrete particle masses and positions as
\begin{equation}\label{eq:general_Replace_density}
\rho_{R}(\bm{x}, z) = \rho_D(\bm{x}, z)  + \sum_{i \in A} \left[\rho_H^i(\bm{x}, z) - \rho_{D}^i(\bm{x}, z)\right],
\end{equation}
where $\rho_D(\bm{x}, z)$ is the density of particles at coordinate $\bm{x}$ and redshift $z$ in a dark matter only simulation. Given some set of coordinates described by $A$, the dark matter mass inside some volume at that location, $\rho_D^i$, can be modified through a replacement with the hydrodynamical density, $\rho_H^i$, which includes any relevant dark matter, gas, stellar, or black hole mass in the volume. The above framework is general such that $A$ can signify any chosen coordinates, but for this work, we will choose $A$ to be a set of halos with masses $M\in[M_a, M_b]$ and radial shell factors $r\in[\alpha_i, \alpha_j]R_{200}$, where $\alpha$ represents some multiple of the radius, $R_{200}$,, where the density is $200\times$ the cosmic density, $\rho_{c}$.  For models where $M_b$ is unbound, i.e. $M_b=\infty$, we will refer to $M_a$ as $M_{\rm min}$, and similarly, when $\alpha_i=0$, we will refer to $\alpha_j$ as $\alpha_{\rm max}$.

To avoid double-counting when halo regions overlap, we define an indicator function as described below such that, 

\begin{equation}\label{eq:Replace}
\begin{split}
    \rho_R(\bm{x},z) = &\rho_D(\bm{x},z) + \\&\sum_{i \in A(M,z)} \theta_i(\bm{x})
\left[\rho_H^i(\bm{x},\alpha) - \rho_D^i(\bm{x},\alpha)\right].
\end{split}
\end{equation}
We assign priority in decreasing order of halo mass; namely, if multiple halos’ Replacement regions contain a given location $\bm{x}$, only the most massive halo performs the Replacement at that location. In more detail, we can think of overlapping halos at some location $\bm{x}$ as those in the set,
\begin{equation}
    O(\bm{x}) = \{j:|\bm{x}-\bm{x_j}|\leq\alpha R^j_{200}\},
\end{equation}
and the halo with priority ordered by mass as $\text{arg max}_{j\in O(\bm{x})}M_j$ such that the winning halo at position $\bm{x}$ has index $i^*$. For scenarios where there is no overlap around any halo, i.e. the particles are unique to a single halo with no others overlapping, the winning halo is undefined by construction, 
\begin{equation}
\begin{split}
    i^*(\bm{x}) = \begin{cases}
    \text{arg max}_{j\in O(\bm{x})}M_j, \ \ &O(\bm{x})\neq \varnothing\\
    \text{undefined},  &O(\bm{x})=\varnothing.
\end{cases}
\end{split}
\end{equation}
Then, the indicator function can determine whether a replacement occurs at location $\bm{x}$ following
\begin{equation}
\begin{split}
        \theta_i(\bm{x}) = \begin{cases}
        1,  \ \ \ \ &i\in O(\bm{x})\,\text{and}\, i=i^*(\bm{x})\\
        0, &\text{Otherwise.}
    \end{cases}
    \end{split}
\end{equation}
Now, if there is no overlap amongst halos, $O(\bm{x}) = i(\bm{x})$ and $i^*(\bm{x}) = i(\bm{x})$, so $\theta_i(\bm{x}) =1$. If there is overlap between multiple halos, then at location $\bm{x}$, the indicator $\theta(\bm{x})=1$  only occurs for the massive halo with $i^*(\bm{x})$.

% To avoid double-counting when halo regions overlap, we assign priority in decreasing order of halo mass. For a given location $\bm{x}$, if multiple halos’ Replacement regions contain $\bm{x}$, only the most massive halo performs the Replacement at that location. Mathematically, we define:

% \begin{equation}\label{eq:Replace}
% m_R(\bm{x},z) = m_D(\bm{x},z) + \sum_{i \in A(M,z)} \theta_i(\bm{x})
% \left[m_H^i(\bm{x},\alpha) - m_D^i(\bm{x},\alpha)\right],
% \end{equation}

% where the indicator function is
% \begin{equation}
% \theta_i(\bm{x}) = \begin{cases}
% 1 & \text{if } |\bm{x}-\bm{x}_i| < \alpha R{200}^i \text{ and } i = \arg\max_j M_j
% \text{ among all } j \text{ with } |\bm{x}-\bm{x}_j| < \alpha R{200}^j, \\
% 0 & \text{otherwise}.
% \end{cases}
% \end{equation}
% This ensures that each spatial point receives replacement from at most one halo and we do not double count during the procedure.

In the limits of $M_{\rm min}\rightarrow 0$ with a large enough $\alpha$ or $\alpha\rightarrow\infty$ with a small enough $M_{\rm min}$, the field becomes purely hydro, and in the opposite limits, the field becomes purely DMO. It can then be thought of as $\rho_R$ interpolating between the DMO and hydrodynamical fields as the mass threshold and radius factor are varied.

For any statistic that we measure from the resulting \textit{Replace} field ($S_R$), one might naively expect that its value is somewhere between the dark matter only ($S_D$) and hydrodynamical versions ($S_H$) of the same statistic. Motivated by this intuition, we define a response fraction,
\begin{equation}\label{eq:cum_resp_func}
F_S(M_{\rm min}, \alpha_{\rm max}) = \dfrac{S_R(M_{\rm min}, \alpha_{\rm max}) - S_D}{S_H-S_D}.
\end{equation}
For a fixed $\alpha_{\rm max}$ and redshift $z$, we can see that $F_S\rightarrow 0$ as $M_{\rm min}\rightarrow \infty$, and $F_S\rightarrow1$ as $M_{\rm min}\rightarrow0$ (so long as $\alpha_{\rm max}$ is sufficiently large). It is important to note that the replacement region is centered on the physical coordinates in the N-body simulation, and the same physical coordinates are excised from the hydro simulation and pasted in. We do not center the replacement region in hydrodynamic simulations on the associated hydrodynamic halos. In this way, the replacement is truly field-level and reflects potential shifts in halo centers between dark matter-only and hydro halos. 

The fraction, $F_S$, is implicitly a function of the chosen statistics. For example, the matter power spectrum, $P(k)$, can be computed for hydro, DMO, and \textit{Replace} fields, yielding a quantity, $F_P(M_{\rm min}, \alpha_{\rm max}; k)$, representing the quantity $F_P$ at each scale $k$. It is often the case that hydrodynamical and DMO statistics are relatively consistent, for example, at large scales in the matter power spectrum. In such a case, the fraction becomes undefined, or if affected by numerical noise, can shoot above or below the 0-1 bound, so we must artificially fix the fraction to some constant, $F_p = c$, or remove such regions from our analysis altogether. In this work, we choose the latter, ignoring regions of $F_S$ where the hydro and DMO differ by $<1\%$.

It should also be noted that the fraction is not necessarily bounded between $0$ and $1$ even if there truly exist significant DMO and hydrodynamical differences. The Replacement in this work does not enforce mass conservation can therefore create a statistic, $S_R$, that is not smoothly interpolated between the DMO and hydrodynamical statistics. For example, when cores of low-mass halos are replaced with their hydrodynamical counterparts, the enclosed mass typically decreases relative to their dark-matter-only counterparts due to feedback-driven mass expulsion. In such a case, we do not capture, or model, the redistribution outside of halo cores, which may be impact some statistics. Instead, $S_R$ isolates the sensitivity of a statistic to the regions where replacements have been made. We make this design choice to avoid conflating physical responses with extra assumptions and approximations for how mass is redistributed. So, the results should be interpreted as responses to perfect matter reconstructions only in the replacement regions. 

The above definitions using $M_{\rm min}$ and $\alpha_{\rm max}$ are inherently cumulative, incorporating all halos above some threshold mass, and replacement regions in spheres according to $\alpha_{\rm max}$. The response fraction then tells us, for a given statistic, how low and how far out we need to go to capture the full baryonic effects on the given statistic.

We can also define an analogous discrete (or commonly referred to in the text as binned) version of Eq.~\eqref{eq:cum_resp_func} where we partition halo masses and radii into finite bins and define the response from a specific mass-bin $[M_a, M_{a+1})$ and radial shell $[\alpha_i R_{200}, \alpha_{i+1} R_{200})$ combination. For brevity, we denote the statistic measured from the \textit{Replace} field in a given bin as, 
\begin{equation}
    S_R^{a,i} \equiv S_R(M_a, M_{a+1}; \alpha_i, \alpha_{i+1}).
\end{equation}
The discrete response fraction in a given mass bin and radial shell is then:
\begin{equation}\label{eq:disc_resp_func}
\begin{split}
\Delta F^{\rm a,i}_{S} =
\dfrac{S_R^{a,i}- S_D}{S_H-S_D},
\end{split}
\end{equation}

If the response to baryons in given mass and radial shells is linear, the cumulative and discrete responses are connected,
\begin{equation}\label{eq:additive_approx}
F_S(M_{\min}, \alpha_{\rm max}) \approx \sum_{\substack{a: M_a \geq M_{\min} \\
i: \alpha_{i+1} \leq \alpha_{\rm max}}}
\Delta F^{\rm a,i}_S,
\end{equation}
where the sum extends over all mass-bin indices $a$ with $M_a \geq M_{\min}$
and all radius bin indices $i$ with $\alpha_{i+1} \leq \alpha_{\rm max}$. Deviations from
this relation then quantifies the level of nonlinearity of the map $S[m]$ with respect to baryonic effects originating in mass bins and radial shells. We can diagnose the level of this non-linearity at each scale of a statistic by defining the fractional non-additivity as:
\begin{equation}\label{eq:non_additivity}
\begin{split}
\epsilon_S(M_{\min}, \alpha_{\rm max}) = 1- \dfrac{
\sum_{\substack{a: M_a \geq M_{\min} \\ i: \alpha_{i+1} \leq \alpha_{\rm max}}}
\Delta F^{\rm a,i}_S}{F_S(M_{\min}, \alpha_{\rm max})}
\end{split}
\end{equation}
Values $|\epsilon_S| \approx 0$ indicate that baryonic impacts on a statistic are separable in mass-radius, while deviations from $0$ indicate nonlinear interactions or correlations between different regions of mass-radius parameter space. It should be emphasized that $\epsilon=0$ does not imply that a statistic itself is linear, though it is a requirement for linear statistics.

A similar nonlinearity has been studied in \citet{Masaki20} using anisotropic separate-universe simulations, where the authors find that non-additive effects from large-scale tidal fields and two-halo terms lead to deviations from linearity. In our context, deviations from $0$ in $\epsilon$ likely reflect contributions from the two-halo term, arising either from replacements of halo exteriors, with neighboring halo fields affected as well, or from the statistic itself being non-linear in mass and radius bins.

Finally, we often summarize a model's performance by taking the signal-to-noise ratio (SNR) weighted mean of $F_S$. We do this by computing,
\begin{equation}
\label{eq:weighted_mean_response}
\begin{split}
&\langle F^m_S \rangle_x = \frac{\sum_x w(x) F^m_S(x)}{\sum_x w(x)}, \\& w(x) = \left(\frac{|S_{\rm H}(x) - S_{\rm D}(x)|}{S_{\rm D}(x)}\right)\left(\dfrac{1}{\sigma_{m,x}^2}\right),
\end{split}
\end{equation}
where the sum extends over all $x$-bins corresponding to a given statistic, $S$, (for example, $k$ for matter power spectrum, or $\ell$ for the angular power spectrum), the model $m$ corresponds to a model configuration in replacement $(M_{\rm min}-\alpha_{\rm max})$ or $(M_a, M_{a+1}; \alpha_i, \alpha_{i+1})$ space. The weights $w(k)$ are SNR-weights, where the ``signal" is the hydro to DMO difference, and the noise is from a given model’s measurement error. This weighting ensures that scales where baryonic effects are largest contribute most to the average, provided the model's measurements have low noise. In contrast, regions where the hydro and DMO models converge, or where noise levels are high, receive negligible weight. This yields a single number summarizing the overall effectiveness of a given model in capturing the baryonic signature on a given statistic.

Throughout the body of the text, we refer to the response and SNR-weighted response fractions defined above, which we emphasize, scales where baryonic effects are strongest, and where the measurements are most precise (lowest $\sigma^2)$. For intuition and completeness, we also present the corresponding unweighted deviation between statistics generated with the most aggressive replace model ($M\geq10^{12}\,h^{-1}\,{\rm M}_\odot$ and $r<5R_{200}$) and full hydrodynamical runs in Appendix~\ref{app:statistica_deviations_from_hydro}

\section{Simulations and Model Creation}
\label{sec:simulations}

\subsection{The IllustrisTNG Suite}

We use the highest-resolution version of the $205\,{\rm Mpc}\,h^{-1}$ box in the IllustrisTNG suite, the TNG300-1 simulation, alongside its dark matter-only counterpart, IllustrisTNG-300-Dark~\citep[][hereafter TNG and TNG-DM, respectively]{Pillepich17, springel18,Nelson2018,naiman18,Marinacci18} run with the adaptive moving-mesh code \texttt{AREPO} \citep{Volker10}. The TNG300-1 simulation spans a periodic volume with a side length of $205\,{\rm Mpc}\,h^{-1}$ and evolves $2500^3$ dark matter and $2500^3$ gas resolution elements, which can become stars and black holes, from redshift $z=127$ to $z=0$.

In practice, the formalism of \S~\ref{sec:formalism} can be applied to any simulation suite containing matched initial conditions between hydrodynamical and dark matter only counterparts. We choose IllustrisTNG for the present work to build upon and extend previous results examining baryonic effects through halo Replacement procedures \citep{Miller-25}, and our previous weak lensing statistic studies \citep{Osato21, leeComparingWeakLensing2022, Lee26}. Understanding the impact of different subgrid physics implementations on the response kernels derived in \S~\ref{sec:formalism} would be valuable, but we leave a systematic comparison across simulation suites (e.g., FLAMINGO \citep{Flamingo-23}, SIMBA \citep{SIMBA}, EAGLE \citep{EAGLE}) for future work.

Both TNG and TNG-DM adopt cosmological parameters consistent with Planck 2015 results: $\Omega_{\Lambda,0}=0.6911$, $\Omega_{m,0}=0.3089$, $\Omega_{b,0}=0.0486$, $\sigma_8=0.8159$, $n_s=0.9667$, and $h=0.6774$ \citep{Ade16}. The TNG simulation incorporates baryonic physics via a subgrid prescription including radiative cooling, star formation, stellar evolution and feedback, supermassive black hole seeding and growth, and AGN feedback \citep{Pillepich-2018}. The model was calibrated to reproduce the galaxy stellar mass function, stellar-to-halo mass relation, and cosmic star formation rate density at $z\lesssim 10$ \citep{Pillepich-2018}.

The mass resolution of TNG300-1 is $m_{\rm gas}=7.44\times 10^6\,h^{-1}\,{\rm M}_\odot$ for initial gas cells and $m_{\rm DM}=3.98\times 10^7\,h^{-1}\,{\rm M}_\odot$ for dark matter particles. The TNG-DM counterpart uses $m_{\rm DM}=4.72\times 10^7\,h^{-1}\,{\rm M}_\odot$, accounting for the cosmic baryon fraction. Both simulations output particle data at 100 snapshots spanning $0 \leq z \leq 20$, with all data publicly available through the IllustrisTNG data release \citep{Nelson2019}.

\subsection{Halo Catalogs and Matching}
Both TNG and TNG-DM employ the Friends-of-Friends (FoF) algorithm with linking length $b=0.2$ to identify dark matter halos, followed by the SUBFIND algorithm \citep{Springel01} to locate gravitationally bound substructures within each FoF group. These catalogs are computed on-the-fly during the simulation and output at each snapshot.

In \S~\ref{sec:BCM}, we compare density profiles from baryon correction models to hydrodynamical profiles from IllustrisTNG. To make such a comparison, we must select a set of DMO halos to be baryonified and have a corresponding hydrodynamic halo counterpart. We accomplish this by constructing catalogs of bijectively matched halos between TNG-DM and TNG. For each halo in TNG-DM, we identify all associated dark matter particles and determine their halo membership in TNG, selecting the TNG halo containing the largest fraction of these particles as the match. We then reverse the process: starting from the matched TNG halo, we verify that it maps back to the original TNG-DM halo. We retain only halo pairs for which $>70\%$ of particles successfully return to their original halo in this bidirectional matching, thereby guaranteeing robust correspondence \citep{Knebe11}. At $z=0$, this matched catalog contains approximately $1.2\times 10^6$ halos with $M_{200} > 10^{12}\,h^{-1}{\rm M}_\odot$ or more than $97\%$ of all halos \citep{Bose2019}.

\subsection{\textit{Replace} Field Construction}
\begin{figure*}[!t]
\centering
\includegraphics[width=\linewidth]{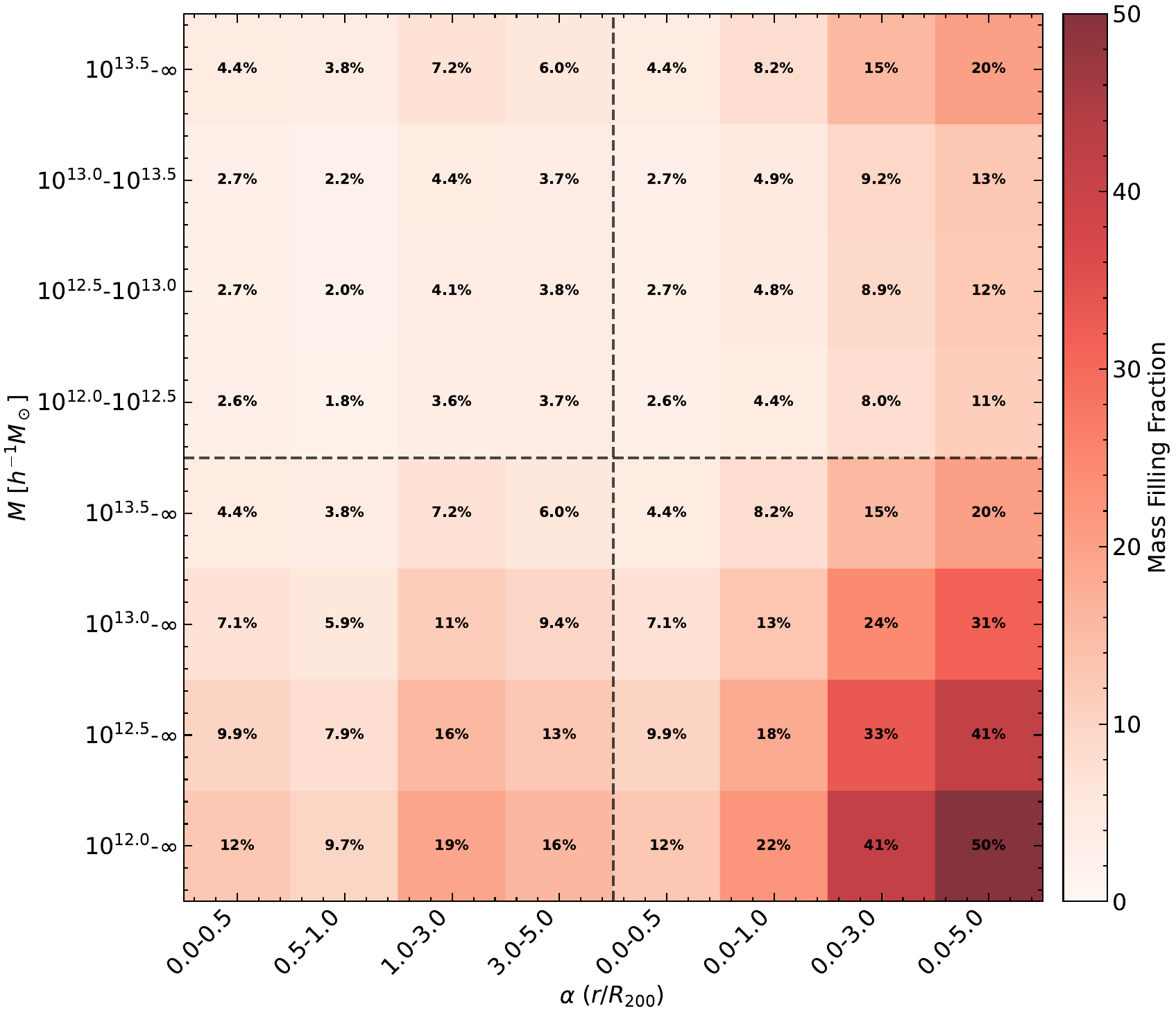}
\caption{Grid showing all \textit{Replace} models constructed in this work. Each tile represents a specific mass-bin radial-shell configuration, with color indicating the mass fraction of the Replacement region within the TNG300-1 box at $z=0$. The upper-left corner shows discrete (single-bin) models, while the lower-right corner shows fully cumulative models. Note that the $\alpha\in[0, 0.5] \cup \log (M/{\rm M}_\odot)\in[13.5, \infty]$ model is repeated in the figure for clarity, as this model is both discrete (it represents a bin) and cumulative (it represents the smallest possible Replacement). Off-diagonal tiles represent mixed configurations (e.g., cumulative in mass, discrete in radius). The most extensive \textit{Replace}—halos with $M_{min}>10^{12} h^{-1} {\rm M}_\odot$ out to $r\leq 5R_{200}$—occupy approximately 50\% of the total mass and are in the lower-right corner.}
\label{fig:models}
\end{figure*}
With the formalism outlined in \S~\ref{sec:formalism}, we construct a suite of \textit{Replace} fields by selectively replacing TNG-DM halo particles with their TNG counterparts at various redshifts. We partition the halo mass function into bins with edges $M_i = [10^{12}, 10^{12.5}, 10^{13}, 10^{13.5}, \infty)\,h^{-1}\,{\rm M}_\odot$ roughly corresponding to Milky Way mass halos, intermediate mass halos, galaxy groups, and massive halos. We choose bins in radius space with shells of factors $\alpha_i = [0, 0.5, 1.0, 3.0, 5.0]R_{200}$ roughly corresponding to the core, one virial radius, the nearby outskirts, and the faraway outskirts. This yields a 4$\times$4 grid in mass-radius space, at each of the 20 snapshots from $z=0.03-2.56$ (chosen for the ray tracing procedure of \S~\ref{sec:raytracing}).

We distinguish between \emph{discrete} models, where halos only within a single mass-radius shell (e.g., $10^{13} \leq M < 10^{13.5}\,h^{-1}{\rm M}_\odot$ and $0.5 < \alpha \leq 1.0$) are replaced, and \emph{cumulative} models, which include all halos above a mass threshold and within a cumulative radius factor (e.g., $M \geq 10^{12}\,h^{-1}{\rm M}_\odot$ and $\alpha \leq 3.0$). Mixed configurations—cumulative in mass but discrete in radius, or vice versa—allow us to isolate the contributions from specific radial shells across the whole mass range, or from specific mass-bins across all radii. Note that the minimum possible Replacement, of $\alpha\in[0, 0.5] \cup \log_{10}(M/{\rm M}_{\odot})\in[13.5, \infty]$ is considered to be both discrete and cumulative according to our above definitions.  Including combinations of both discrete in radius/mass and cumulative in radius/mass configurations, this procedure yields 64 \textit{Replace} field variants.

In Fig.~\ref{fig:models}, we show a grid representing each \textit{Replace} field model. The upper-left corner represents discrete models, whereas the lower-right corner shows purely cumulative models. The off-diagonal elements show combinations of cumulative mass/radius bins with discrete radius/mass bins. The color of each tile indicates the mass fraction of the replacement region within the TNG300-1 box at $z=0$. We compute the mass fraction by summing the masses of all dark matter particles within a given \textit{Replace} region and dividing by the total mass of the simulation box. The most extensive cumulative configuration, replacing halos with $M\geq10^{12}\,h^{-1}{\rm M}_\odot$ out to $5R_{200}$, follows this procedure and removes approximately 50\% of the N-body simulation mass, replacing it with all hydrodynamic particles, including dark matter, gas, and stellar particles. Lower-mass halos and smaller radii naturally occupy smaller mass fractions, with the discrete bin models each affecting $\lesssim 10\%$ of the mass.

One consideration when performing the Replacement is the effect of sharp boundaries induced by a strict radial cutoff. Another consideration is the halo centering during the replacement. In this work, we take the positions of dark matter-only halos to be the halo centers in both the dmo and hydro simulations. However, there is an offset between the dmo and hydro halos, which may lead to offset replacement regions. To probe both of these choices, in Appendix~\ref{appendix:cutoff}, we explore the effects of a hard Replacement vs. a softer weighted ramping. We find that the cutoff effects are less than 1\% across all relevant scales in this work and therefore retain the hard cutoff and halo centering choices for clarity in interpreting our results. 

\subsection{Ray-Tracing Pipeline}\label{sec:raytracing}

We employ a multi-lens-plane ray-tracing approach to generate weak lensing convergence from the \textit{Replace} fields, following the methodology of \citet{jain2000, Hilbert09, Petri16a, Osato21, leeComparingWeakLensing2022, Ferlito24}. We briefly summarize the key steps here and refer the reader to these works for full implementation details.

Light rays are traced backwards from an observer at $z=0$ to a source redshift $z_s$. To construct light cones spanning comoving distances up to $\sim 4000\,h^{-1}{\rm Mpc}$ (corresponding to $z_s \approx 2.56$), we stack \textit{Replace} snapshots along the line of sight at fixed intervals equal to the box size, $205\,h^{-1}{\rm Mpc}$. This requires 20 snapshots in total. Following \citet{Osato21}, we select the TNG snapshots closest to the centres of these 20 equally-spaced comoving distance intervals. The specific snapshot numbers, redshifts, and comoving distances are listed in Table~1 of \citet{Osato21}.

Each snapshot is then split into two planes spaced at regular intervals of half the TNG300-1 comoving box size, $\Delta \chi = 102.5\,h^{-1}{\rm Mpc}$. At each plane, we project the 3D mass field $\rho(\bm{x},z)$ onto a 2D surface density $\Sigma(\bm{\theta},\chi)$ by integrating along the line of sight over the plane thickness:
\begin{equation}\label{eq:delta_plane}
\Sigma(\bm{\theta}, \chi_i) = \int_{\chi_i - \Delta\chi/2}^{\chi_i + \Delta\chi/2} \rho(\chi \bm{\theta}, \chi), d\chi,
\end{equation}
where $\bm{\theta}$ is the angular position on the sky and $\chi$ is comoving distance. Following the weak lensing suite of \citet{Osato21} (henceforth called $\kappa$TNG), we perform ray tracing in a $5\times 5\deg^2$ light cone. The last 10 snapshots ($z \gtrsim 0.8$) are replicated 4 times in the transverse direction to achieve the desired angular coverage at high redshift \citep{Osato21}.

The deflection angle $\bm{\alpha}$ accumulated by a light ray passing through lens plane $i$ is computed from the 2D gravitational potential via
\begin{equation}
\bm{\alpha}_i(\bm{\theta}) = \frac{2}{c^2} \int \nabla{\perp} \Phi(\bm{\theta}', \chi_i) \frac{d^2\theta’}{|\bm{\theta} - \bm{\theta}'|},
\end{equation}
where $\Phi$ is related to $\Sigma$ through the 2D Poisson equation. In practice, we solve this in Fourier space using Fast Fourier Transforms (FFT) on a $4096^2$ grid spanning the light cone. Dark matter, gas, and stellar particles (or, for \textit{Replace} fields, the hybrid mass distribution) are assigned to the grid using a triangular-shaped cloud (TSC) scheme.

The convergence $\kappa$ and shear $\gamma$ fields are derived from second derivatives of the lensing potential associated with the Jacobian of deflections. The final convergence and shear maps are output on a $1024^2$ pixel grid with pixel size $0.29\,{\rm arcmin}$, consistent with the resolution used in the public $\kappa$TNG suite.

To suppress sample variance, we generate multiple pseudo-independent realizations of each \textit{Replace} model by exploiting the periodic boundary conditions and the fact that the light cone does not cover the entire simulation box. For each model, we translate all particles by a random offset along each of the three axes, rotate the snapshot by 0, 90, 180, or 270 degrees around each axis, and apply a random flip along any axis. This procedure is applied 20 times to each simulation snapshot before projecting onto 2D density planes, yielding 20 distinct sets. To generate convergence maps, we randomly translate and rotate the above set of 2D density planes, yielding 100 convergence maps per density-plane set. Altogether, we end up with 2,000 convergence maps for each of the \textit{Replace} models, as well as for TNG and TNG-DM. It is important to note that these convergence maps are pseudo-independent, meaning they are random realizations of convergence maps constructed from a simulation with the same initial conditions. As shown in \citet{Petri16}, the above method can generate up to 10,000 statistically independent realizations.

The above procedure then yields a total of $5.28$ million convergence maps, $\kappa$, across all models and 40 source redshifts. We make this dataset publicly available for community use and exploration.\footnote{Data release details and access instructions are available at https://columbialensing.github.io/}

\subsection{Statistical responsess}\label{sec:statistics}

We apply the replacement procedure of \S~\ref{sec:formalism} to various statistics in this work and discuss their computation here.  For each of the 20 TNG300-1 snapshots spanning redshifts $0.03 \leq z \leq 2.56$, particles are assigned to two $4096^2$ pixel grids following the rotation and translation procedure described in the previous section. We compute the overdensity field as
\begin{equation}
\delta_i(\bm{x}, z) = \frac{\rho_i(\bm{x}, z) - \bar{\rho}_i(z)}{\bar{\rho}_i(z)},
\end{equation}
where $\bar{\rho}_i(z)$ is the mean density of model $i$ at redshift $z$. The matter power spectrum is then computed via
\begin{equation}\label{eq:power_spec}
P_i(k, z) = \langle |\tilde{\delta}_i(\bm{k}, z)|^2 \rangle,
\end{equation}
where $\tilde{\delta}_i$ is the Fourier transform of $\delta_i$ and angle brackets denote averaging over Fourier modes in $2,047$ logarithmic $k$-bins between $k=0.037\rightarrow62.7\,h\,{\rm Mpc}^{-1}$ and over the pseudo-independent realizations.

For weak lensing statistics, we perform the above procedure to generate the set of convergence maps for each \textit{Replace} model. We  then compute for each convergence map the angular power spectrum,
\begin{equation}\label{eq:C_l}
C_{\kappa}(\ell_i) = \langle|\tilde{\kappa}(\bm{\ell})|^2\rangle_{\bm{\ell} \in [\ell_i^{\rm min}, \ell_i^{\rm max}]},
\end{equation}
where $\tilde{\kappa}(\bm{\ell})$ is the two-dimensional Fourier transform of the convergence field $\kappa(\bm{\theta})$, and the angle brackets denote averaging over all Fourier modes $\bm{\ell}$ within the multipole bin $[\ell_i^{\rm min}, \ell_i^{\rm max}]$.  

For non-Gaussian statistics, we focus on four statistics computed using the \emph{LensTools} package \citep{Petri16a}\footnote{\url{lenstools.readthedocs.io}} from the convergence maps:
\begin{enumerate}
\item \textbf{Peak counts} $N_p(\nu)$: The number of local maxima in the convergence field, defined as pixels with $\nu$ with greater values than their $8$ nearest neighbors, as a function of signal-to-noise ratio $\nu = \kappa/\sigma_\kappa$, where $\sigma_\kappa = \sqrt{\langle \kappa^2 \rangle}$ is the root-mean-square convergence \citep{Maturi+2010}.
\item \textbf{Minimum counts} $N_{\rm m}(\nu)$: The number of local minima in the convergence field with $\nu$ values smaller than their $8$-nearest neighbors \citep{coulton20}.
\item \textbf{One-point PDF} $p(\nu)$: The probability distribution of convergence values representing the full histogram of $\nu$ values \citep{thiele_accurate_2020}.
\item \textbf{Minkowski functionals} (MFs): Three descriptors of convergence excursion sets \citep{Kratochvil+2012}—area $V_0(\nu)$, boundary length $V_1(\nu)$, and genus $V_2(\nu)$—that encode shape information.
\end{enumerate} 
These statistics complement the angular power spectrum by accessing non-Gaussian features of the matter distribution, which have been shown to hold a wealth of cosmological information \citep{Zurcher2021}

% This yields angular power spectra for each of the 2,000 pseudo-independent maps across all \textit{Replace} models and source redshifts. We compute the same quantities for the non-\textit{Replace} models (i.e., pure TNG-hydro and TNG-DMO), allowing us to define the cumulative response fraction for the convergence power spectrum following Eq.~\eqref{eq:cum_resp_func}:
% \begin{equation}\label{eq:F_cl}
% F_{C_\ell}(M_{\min}, \alpha_{\rm max}; \ell, z_s) = \frac{\langle C_\ell^{\rm R}(M_{\min}, \alpha_{\rm max}; z_s) \rangle - \langle C_\ell^{\rm DMO}(z_s) \rangle}{\langle C_\ell^{\rm H}(z_s) \rangle - \langle C_\ell^{\rm DMO}(z_s) \rangle},
% \end{equation}
% where the angle brackets denote averaging the measured statistic over the 2,000 map realizations, and an equivalent statistic can be measured for discrete models with mass bins and radial shells $(M_a, M_{a+1}; \alpha_i, \alpha_{i+1})$.

\section{The Baryonic Response of the Matter Field}
\label{sec:results_matter}

We begin our analysis with the matter power spectrum. Our approach proceeds similarly to the halo replacement analysis of \citet{Miller-25} and extends it across a systematic grid of mass bins, radial shells, and redshifts. We describe all results within the response formalism developed in \S~\ref{sec:formalism}.

\subsection{Matter Power Spectrum}

\begin{figure*}[!t]
\centering
\includegraphics[width=\linewidth]{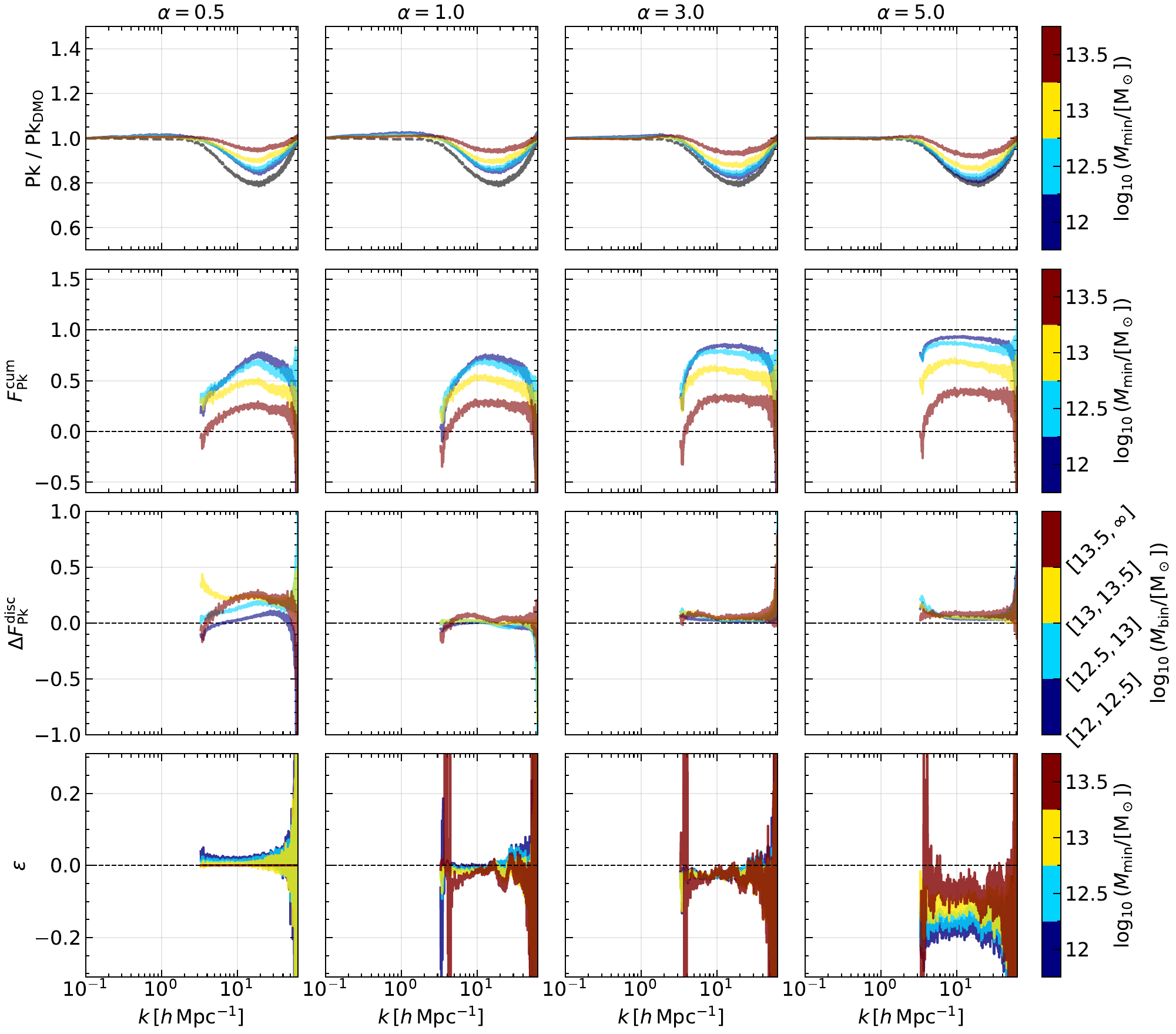}
\caption{Decomposing the matter power spectrum suppression by mass and radius factor at $z=0.03$. Each column corresponds to a radius factor $\alpha$ (indicated at top), with curves colored by mass thresholds or bins. \emph{Top row:} Ratio of \textit{Replace} to DMO power spectra $P_{\rm R}(k)/P_{\rm D}(k)$ for cumulative models. The black curve shows the target Hydro-to-DMO ratio $P_{\rm H}(k)/P_{\rm D}(k)$. \emph{Second row:} Cumulative response fractions $F_P(M_{\min}, \alpha_{\rm max})$ from Eq.~\eqref{eq:cum_resp_func}. Values of $F=0$ are consistent with DMO power spectra, while $F=1$ corresponds to matching the hydrodynamical TNG exactly. Values outside $[0,1]$ indicate replacement artifacts. \emph{Third row:} Discrete bin responses $\Delta F_P^{\rm a,i}$ from Eq.~\eqref{eq:disc_resp_func}, showing contributions from individual mass-radius shells. \emph{Bottom row:} Non-additivity metric $\epsilon(M_{\min}, \alpha; k)$ from Eq.~\eqref{eq:non_additivity}, measuring the degree to which discrete bins sum linearly to cumulative models. Values near $\epsilon = 0$ indicate linear additivity while deviations indicate non-linear coupling between mass-radius bins.}
\label{fig:suppression_and_fraction}
\end{figure*}

We replace the generic statistic subscript $S$ with $P$ for the matter power spectrum of Eq.~\ref{eq:power_spec}. We compute responses for all models in Fig.~\ref{fig:models}. Fig.~\ref{fig:suppression_and_fraction} presents a comprehensive decomposition of power spectrum suppression by mass and radius at $z=0.03$ (snapshot 96). The top row shows the power spectrum suppression, $P_{\rm R}(k)/P_{\rm D}(k)$ for cumulative \textit{Replace} models compared to the full hydrodynamical target $P_{\rm H}(k)/P_{\rm D}(k)$ (black curve).

We find that even the most aggressive \textit{Replace} model ($M_{\rm min}=10^{12}\,h^{-1}{\rm M}_\odot$ out to $\alpha_{\rm max}=5$) cannot fully reproduce the true hydrodynamical suppression. Instead, $\sim 90\%$ of the total hydrodynamical suppression is captured at $3 \lesssim k \lesssim 63\,h\,{\rm Mpc}^{-1}$, consistent with the findings of \citet{Miller-25}. As we extend the radial range used in \citet{Miller-25} from $\alpha_{\rm max}=2\rightarrow5$, we still do not fully  capture the suppression. This implies that a non-negligible fraction of the baryonic signal impacting the power spectrum is due to halos below our mass limit ($M_{\rm min} = 10^{12}\,h^{-1}{\rm M}_\odot$) and/or gas at radii beyond $\alpha_{\rm max}=5$.

We also see that models with small radius factors exhibit spurious large-scale power enhancement, as also reported by \citet{Miller-25}. Configurations with $\alpha_{\rm max} = 0.5$ or $1$ systematically show $P_{\rm R}/P_{\rm D} > 1$ at $k < 1\,h\,{\rm Mpc}^{-1}$. This counterintuitive behavior arises from mass-conservation violations: replacing only halo cores with their lower-mass hydrodynamical counterparts removes mass from high-density peaks without compensating in the outskirts, thereby artificially reducing the mean density $\bar{\rho}$ and inflating the overdensity $\delta = (\rho - \bar{\rho})/\bar{\rho}$ on large scales \citep{Miller-25}. One note is that this large-scale discrepancy might be eliminated if $\bar{\rho}$ is set to a single value, such as the mean of the dark matter-only map, so that overdensities are always relative to a given mean. We choose not to follow this, as it would confuse the interpretation of the \textit{Replace} model at all scales, whereas our analysis already removes large-scale artifacts where the hydro and DMO power spectra are less than $1\%$. The large scale artifact also seems to disappear when $\alpha_{\rm max} \geq 3$, indicating that at roughly $r\gtrsim 3R_{200}$, dark matter only and hydrodynamical halos have roughly equal masses, a result which is consistent with the findings of \citet{Miller-25} and \citet{Ayromlou23}, who investigate baryon closure radii in the IllustrisTNG simulations. 

The second row in Fig.~\ref{fig:suppression_and_fraction} recasts these results as cumulative response fractions $F_P(M_{\min}, \alpha_{\rm max}; k)$ following Eq.~\ref{eq:cum_resp_func}. By construction, $F_P = 0$ indicates the \textit{Replace} model is indistinguishable from DMO, while $F_P = 1$ implies full recovery of the hydrodynamical suppression. Values outside $[0,1]$—clearly visible at high $M_{\min}$ across all $\alpha_{\rm max}$—directly flag mass-conservation artifacts as discussed in the previous paragraph. Deviations from the bounds in $F$ then indicate a systematic error and can be used as a diagnostic. The cumulative responses highlights the relative importance of baryons on the power spectrum in certain mass and radius bins, and how these effects saturate: galaxy groups and clusters with $M \geq 10^{13}\,h^{-1}{\rm M}_\odot$ Replaced out to $5R_{200}$ alone account for $F_P \sim 0.60$ (60\% of the total baryonic effects on the matter power spectrum) across $3 \lesssim k \lesssim 63\,h\,{\rm Mpc}^{-1}$. Including lower-mass halos down to $M_{\min} = 10^{12.5}\,h^{-1}\,{\rm M}_\odot$ increases this to $F_P \sim 0.80$ at $k \lesssim 50\,h\,{\rm Mpc}^{-1}$. Further extending the mass threshold to $10^{12}\,h^{-1}\,{\rm M}_\odot$ provides additional gains, pushing $F_P$ to $\sim 0.9$ at nearly all applicable scales but never fully reaching a perfect score of $F=1.0$.

The third row in Fig.~\ref{fig:suppression_and_fraction} explores the contributions of individual mass bin-radius shell \textit{Replace} models via the discrete response of Eq.~\eqref{eq:disc_resp_func}. This shows a more detailed spatial origin of baryonic effects on the power spectrum. The cores of halos ($\alpha \in [0, 0.5)$, leftmost panel) host the most significant effects: baryons in group-scale halos ($10^{13} \leq M\,[h^{-1}\,{\rm M}_\odot] < 10^{13.5}$ dominate suppression at intermediate scales $3 \lesssim k \lesssim 10\,h\,{\rm Mpc}^{-1}$ with $\Delta F_P^{\rm bin} \sim 0.25$, while massive halos ($M \geq 10^{13.5}\,h^{-1}{\rm M}_\odot$) take over on smaller scales $k \gtrsim 3\,h\,{\rm Mpc}^{-1}$. Moving to outer shells ($\alpha \in[0.5,1)$ and beyond), the contribution pattern shifts: galaxy clusters maintain a nearly constant $\Delta F_P^{\rm bin} \sim 0.05$–$0.1$ across all outer radial bins, consistent with AGN feedback redistributing gas uniformly to $r \sim 2$–$3R_{200}$ in massive systems \citep{Ayromlou23}. Lower-mass halos ($M < 10^{13}\,h^{-1}{\rm M}_\odot$) exhibit weaker radial dependence overall.

The bottom row in Fig.~\ref{fig:suppression_and_fraction} quantifies the degree of non-linearity to mass bins and radial shells in the response using Eq.~\eqref{eq:non_additivity}. Values near $\epsilon = 0$ indicate that discrete bins sum linearly to reproduce cumulative models while deviations signal non-additive coupling in mass-radius space. While most configurations exhibit relative additivity ($|\epsilon| \lesssim 0.05$), the highest radius factor $\alpha_{\rm max} = 5$ shows significant non-linearity for all mass thresholds, with low-mass halos displaying $|\epsilon| \sim 0.1$–$0.2$. This breakdown likely arises from contamination by the two-halo term. At large radii ($r \gtrsim 3R_{200}$), the Replacement procedure includes particles associated with correlated environments of neighboring halos, altering the two-halo term. Replacing the outskirts of a single halo then affects not only its own contribution to $P(k)$ but also its correlation with nearby halos, breaking additivity.  It should be emphasized that the non-linearity metric does not refer to the nonlinear mapping between the density field and the statistic itself. Instead, it diagnoses non-linearity of a statistic between radius shells, across halo mass bins, or both. A linear statistic is a sufficient but not necessary condition for $\epsilon=0$.

\begin{figure*}[!t]
\centering
\includegraphics[width=\linewidth]{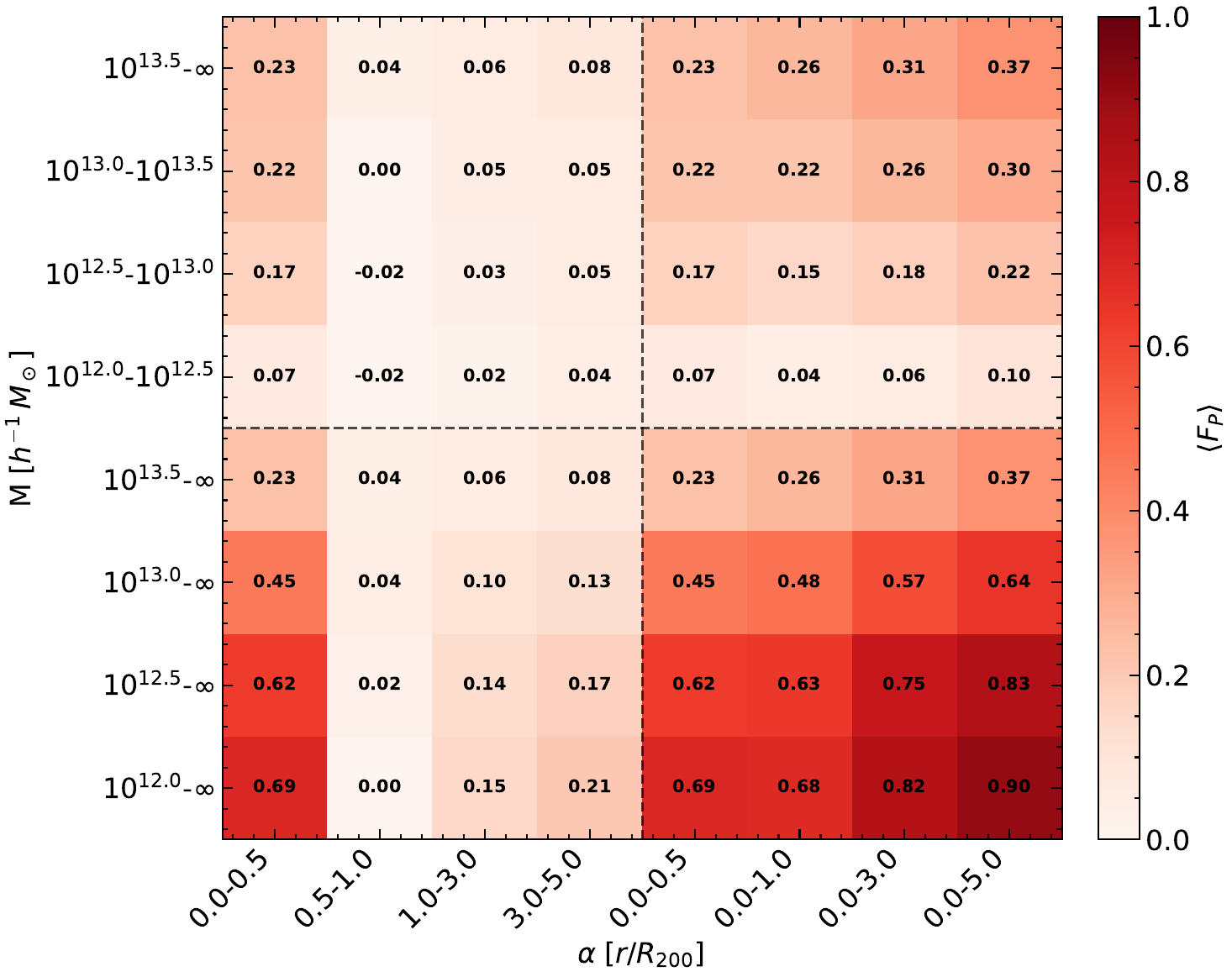}
\caption{Weighted mean power spectrum responses at $z=0.03$, summarizing Fig.~\ref{fig:suppression_and_fraction} into single numbers per model. The structure follows the same as Fig.~\ref{fig:models}. Rows correspond to $M$ bins, columns to radial shells, with the top left corner showing the discrete models, and the bottom right showing the cumulative models. To achieve $\sim 90\%$ fidelity ($\langle F_P \rangle \gtrsim 0.9$, bottom-right cell), one must correctly model the field level of all halos above $10^{12}\,h^{-1}{\rm M}_\odot$ out to at least $5R_{200}$. The brightest cell for the discrete models (dark red, $\langle \Delta F_P^{\rm a,i} \rangle \sim 0.23$) happens to also be a cumulative model representing the minimal possible replacement at $(M_{\rm min},\alpha_{\rm max} ) = ([10^{13.5}, \infty]\,h^{-1}{\rm M}_\odot, [0, 0.5R_{200}))$, confirming that massive halo cores dominate the baryonic response on $P(k)$. The outermost shell ($\alpha\in[3, 5)$) exhibits $F\sim 0.04$–$0.08$ for all mass-bins, likely due to the two-halo term and extended gas redistribution contributions.}
\label{fig:response}
\end{figure*}

To synthesize the scale-dependent information in Fig.~\ref{fig:suppression_and_fraction} into a single metric, we compute the SNR-weighted mean fractional response of Eq.~\eqref{eq:weighted_mean_response}. Fig.~\ref{fig:response} presents $\langle F_P \rangle_k$ at $z=0.03$ as a heatmap over the model space in mass and radius and provides a quick-reference guide to the results of Fig.~\ref{fig:suppression_and_fraction}. To capture $\sim 90\%$ of the baryonic effects on the power spectrum ($\langle F_P \rangle_k \gtrsim 0.9$), one must correctly reconstruct the density distribution of \emph{all} halos above $10^{12}\,h^{-1}{\rm M}_\odot$ out to at least $5R_{200}$ (bottom rightmost corner). More modest goals can be met with less aggressive requirements, such as $\sim65\%$ if only considering massive halos with $M \geq 10^{13} h^{-1} {\rm M}_\odot$.

The upper left corner of Fig.~\ref{fig:response} shows responses for discrete mass-radius shells, revealing the detailed spatial origin of the $P(k)$ suppression. The brightest cell (dark red, $\langle \Delta F_P^{\rm bin} \rangle_k \sim 0.23$) appears at $(M_{\rm min}, \alpha_{\rm max}) = ([10^{13.5}, \infty)\,, [0, 0.5))$, which also corresponds with the minimum replacement model, showing that the majority of baryonic effects on the power spectrum occur in cluster-scale halo cores. The regions between the core and the virial radius show relatively little sensitivity to the baryonic modifications. Still, the exterior shells ($\alpha \in [3, 5]$) show a resurgence to $\langle \Delta F_P^{\rm bin} \rangle_k \sim 0.05$–$0.1$ for all mass-bins. 
%This non-monotonic radial profile reflects extended gas redistribution plus two-halo correlations.
Because the mass and radial bins are approximately independent (except at $\alpha = 5$ where non-additivity emerges, Fig.~\ref{fig:suppression_and_fraction} bottom row), summing across rows or down columns in the upper left corner discrete models roughly reconstructs the cumulative heatmap in the bottom right corner cumulative models, providing a useful consistency check.

It is important to caution the reader’s interpretation of Fig.~\ref{fig:suppression_and_fraction} and Fig.~\ref{fig:response}. First, while the cores of massive halos dominate the response—with the $\alpha\in[0,0.5]$, $M\geq10^{13.5}\,h^{-1}{\rm M}_\odot$ shell alone contributing $\Delta F_P \sim 0.22$, we cannot conclude that precisely modeling cores is sufficient for reproducing the matter power spectrum. The remaining shells jointly contribute $\sim 70\%$ of the response. The cores are necessary but not sufficient.

More fundamentally, it is essential now to distinguish between where baryonic feedback physically acts and the spatial regions to which a statistic is sensitive in the presence of baryons. Our response functions measure the latter, not the former. In the $\alpha_{\rm max}=0.5$, $M_{\rm min}=10^{13.5}\,h^{-1}\,M_\odot$ model (the top left-most corner) of Fig.~\ref{fig:response} it does not imply that $23\%$ of the baryonic mass redistribution occurs within halo cores. Instead, it indicates that $23\%$ of the statistical signature on $P(k)$ originates from differences in the core density profiles between hydrodynamical and dark-matter-only runs.

\subsection{Redshift Evolution of the Power Spectrum Response}

\begin{figure*}[!t]
\centering
\includegraphics[width=\linewidth]{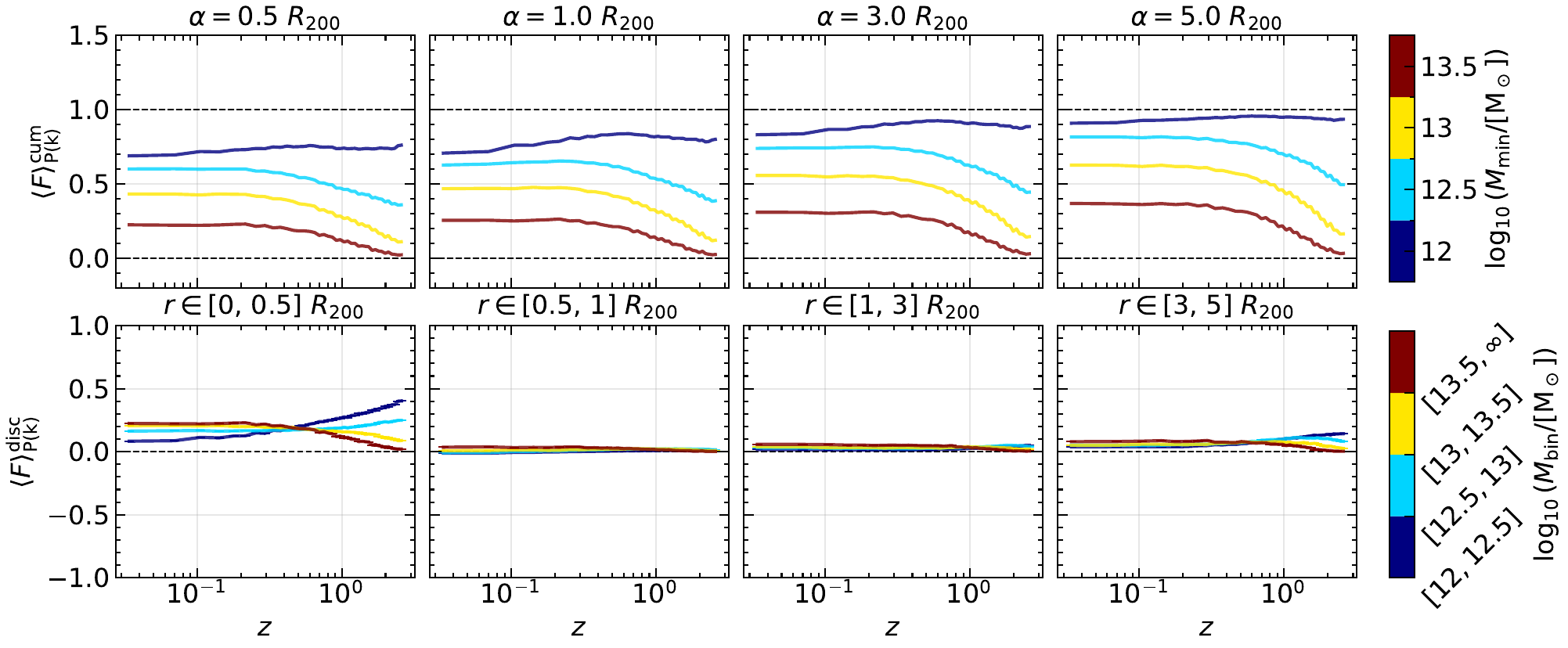}
\caption{Redshift evolution of weighted mean power spectrum responses $\langle F_P \rangle(z)$ following Eq.~\eqref{eq:weighted_mean_response} from $z=2.56$ to $z=0.03$ for cumulative (top panels) and discrete (bottom panels) models. Baryonic effects strengthen toward low redshift for all models with $M\geq 12.5\,h^{-1}{\rm M}_\odot$. At low redshift ($z \lesssim 0.5$), massive halos ($M \geq 10^{13.5}\,h^{-1}{\rm M}_\odot$) dominate the response, whereas at high redshift ($z \gtrsim 0.5$), low mass halos ($M \sim 10^{12}$–$10^{12.5}\,h^{-1}{\rm M}_\odot$) become the dominant source of baryonic effects (bottom row left panel). Outer shells ($\alpha > 0.5$) exhibit weaker redshift evolution, with $\langle \Delta F_P^{\rm bin} \rangle$ remaining roughly constant at $\sim 0.03$–$0.08$ across all $z$.}
\label{fig:power_spectrum_z_evolution}
\end{figure*}

For observables that probe the matter distribution across cosmic time, such as weak gravitational lensing, understanding the redshift dependence of baryonic effects on statistics is crucial. Weak lensing, in particular, represents a projected matter density along a line of sight from an observer to redshift $z$, highlighting the importance of understanding baryonic signatures in statistics as a function of redshift.

Fig.~\ref{fig:power_spectrum_z_evolution} shows the evolution of $\langle F_P \rangle_k$ from $z=2.56$ to $z=0.03$ for all cumulative (top panels) and discrete (bottom panels) models. One notable effect in both the cumulative and discrete models is that baryonic imprints on statistics strengthen toward lower redshifts for all thresholds except the lowest mass threshold. This reflects two effects. One is the increasing abundance of higher-mass halos at lower redshifts, and the other is the growth of feedback-driven modifications as halo assembly proceeds and AGN/stellar feedback intensify. The steepest evolution occurs between $z \sim 2$ and $z \sim 0.3$, coinciding with the epoch of peak star formation and black-hole accretion in the TNG model \citep{Joo2023}. In the discrete models we can see there is a mass-dependent inversion, where at lower redshift ($z \lesssim 0.5$), massive halos ($M \geq 10^{13.5}\,h^{-1}{\rm M}_\odot$) dominate the core response, whereas at higher redshift ($z \gtrsim 0.5$), lower-mass halos ($M \sim 10^{12}$–$10^{12.5}\,h^{-1}{\rm M}_\odot$) become more critical, surpassing clusters.

% The bottom panels decompose this evolution into discrete mass-radius shells showing a similar trend. Halo cores ($\alpha < 0.5$, leftmost bottom panel) show the most substantial baryonic imprint evolution. Crucially, there is a mass-dependent inversion, where at lower redshift ($z \lesssim 0.5$), massive halos ($M \geq 10^{13.5}\,h^{-1}{\rm M}_\odot$) dominate the core response, whereas at higher redshift ($z \gtrsim 0.5$), lower-mass halos ($M \sim 10^{12}$–$10^{12.5}\,h^{-1}{\rm M}_\odot$) become more critical, surpassing clusters.

In outer shells ($\alpha_{\rm max} > 0.5$), the redshift evolution weakens considerably. Contributions from $0.5R_{200} \leq r < R_{200}$ and $R_{200} \leq r < 3R_{200}$ remain roughly constant at $\langle \Delta F_P^{\rm bin} \rangle_k \sim 0.05$ across all redshifts. The weak redshift evolution of outer shells has important implications for weak lensing observables. The rapidly evolving core signal should produce detectable redshift-dependent features in lensing statistics, particularly for observables sensitive to both high and low-mass halos. 

\section{The Baryonic Response of Weak Lensing Observables}
\label{sec:results_WL}
The previous section demonstrated the response of the matter power spectrum to baryonic effects, extending the halo Replacement analysis of \citet{Miller-25} using the response formalism of \S~\ref{sec:formalism}. This allowed a systematic quantification of the importance of baryon physics across halo mass and radius at various redshifts. We now apply the same framework to weak gravitational lensing statistics.

\subsection{Angular Power Spectrum}
\begin{figure*}[!t]
\centering
\includegraphics[width=\linewidth]{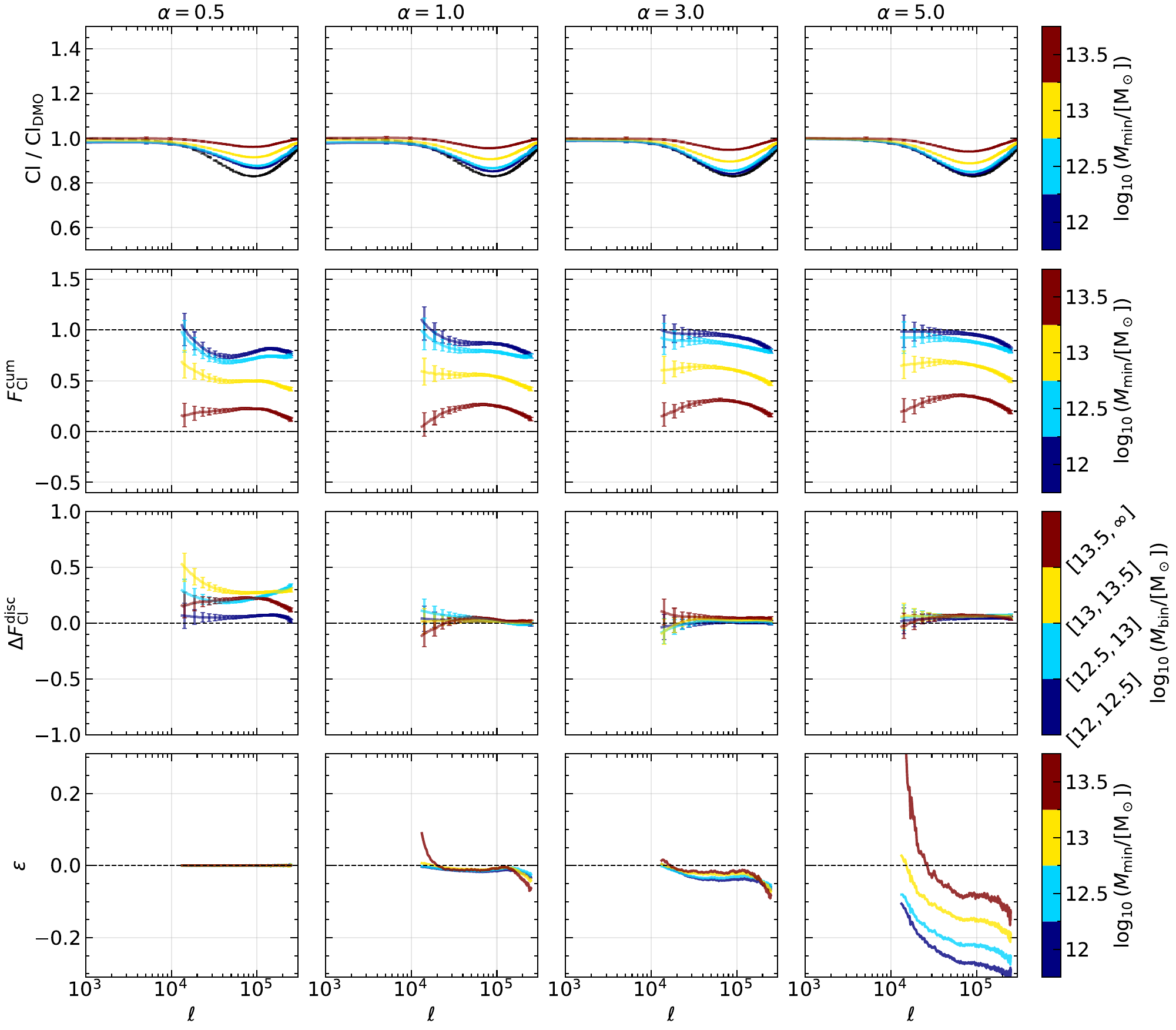}
\caption{Baryonic response of the weak lensing angular power spectrum at galaxy source redshift $z_s = 1.03$. The format of the panels follows the identical structure as Fig.~\ref{fig:suppression_and_fraction}
}
\label{fig:C_ell_scale_dependent_response_z1}
\end{figure*}
\begin{figure*}[!t]
\centering
\includegraphics[width=\linewidth]{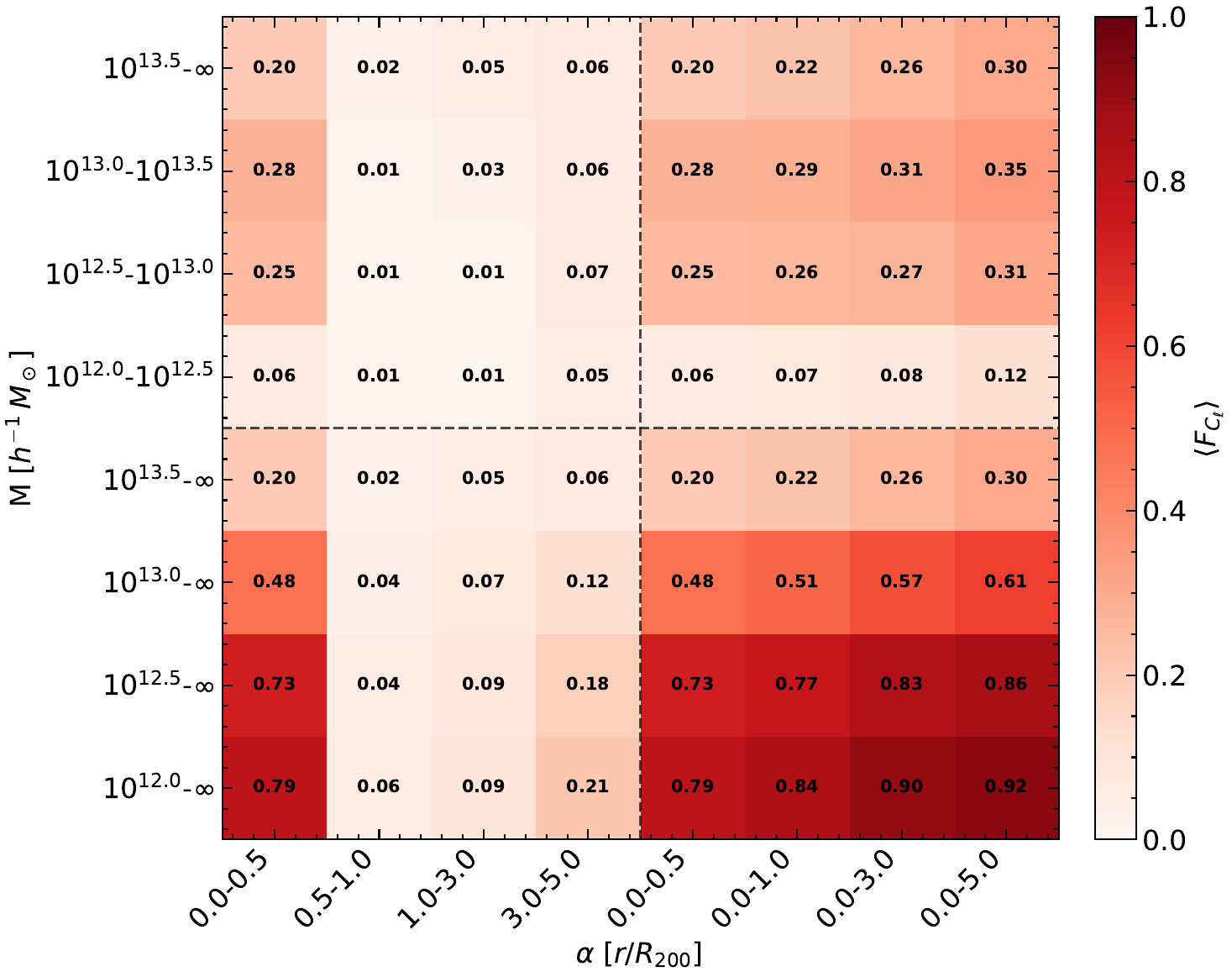}
\caption{Multipole-averaged baryonic response to the weak lensing angular power spectrum at source redshift $z_s \approx 1.0$, following the same format as Fig.~\ref{fig:response}.}
\label{fig:C_ell_cumulative_summary_z1}
\end{figure*}
% \begin{figure}
% \centering
% \includegraphics[width=\linewidth]{imgs/Cl_z_evolution.pdf}
% \caption{Redshift evolution of multipole-averaged angular power spectrum responses from galaxy source redshift $z_s = 0.03$ to $z_s = 2.57$.
% The layout mirrors Fig.~\ref{fig:power_spectrum_z_evolution}.}
% \label{fig:Cl_z_evolution}
% \end{figure}

We begin with the Gaussian statistics of the weak lensing field:

Fig.~\ref{fig:C_ell_scale_dependent_response_z1} is analogous to Fig.~\ref{fig:suppression_and_fraction} and shows the convergence power spectrum suppression (top row), cumulative response fractions (second row), discrete response fractions (third row), and non-additivity statistic (fourth row) at source redshift $z_s = 1.03$. The black curve in the top row shows the complete hydrodynamical suppression $C_\ell^{\rm H}/C_\ell^{\rm DMO}$.

Similar to the matter power spectrum (Fig.~\ref{fig:suppression_and_fraction}), the most aggressive Replacement—all halos with $M \geq 10^{12}\,h^{-1}{\rm M}_\odot$ out to $5R_{200}$—almost entirely reproduces the hydrodynamical suppression across all multipoles but not fully, achieving $F_{C_\ell} \gtrsim 0.95$ at $10^4\lesssim \ell \lesssim 3\times10^5$. However, there is a steep fall off in $F_{C_\ell}$ at smaller scales at high significance.

Models with small radius factors ($\alpha_{\rm max} = 0.5$ or $1$) and low mass show suppression at large scales ($\ell \lesssim 500$), and $C_\ell^{\rm R}/C_\ell^{\rm DMO}$ to exceed the bounds of $0$ and $1$. This occurs in regions where the true baryonic effect is already quite small, and the values of $F$ are already masked in the analysis, though it is interesting to note that this artifact is qualitatively inverse to that seen in the matter power spectrum of Fig.~\ref{fig:response}. This is likely due to the interplay between violations of mass conservation in replacement regions and the weak lensing kernel smoothing these effects in redshift, though a deeper investigation is left to future work.

In the third row of Fig.~\ref{fig:C_ell_scale_dependent_response_z1}, we show the discrete response for the mass and radial bins. Similar to the matter power spectrum, we see the dominant impact from lower mass galaxy groups of $10^{13} \leq M < 10^{13.5}\,h^{-1}\,{\rm M}_\odot$. But, unlike the matter power spectrum, this mass regime and in the centers of halos appears to dominate across virtually all $\ell$. While the response to baryons in $C_\ell$ for massive halos increases at smaller scales, it never outweighs the response for lower mass galaxy groups. At the smallest scales, the lower mass $(10^{12.5}\leq M < 10^{13}\,h^{-1}\,{\rm M}_\odot)$ halos begin to dominate the response and, outside of the halo core, the lower mass halos dominate the baryonic response out to $\alpha =1$. Though for the exteriors of halos $(3 < \alpha \leq 5)$, baryonic effects on $C_\ell$ are seemingly purely sourced by clusters, causing an overall 10\% increase on large to intermediate scales.

The bottom row of Fig.~\ref{fig:C_ell_scale_dependent_response_z1} shows the non-linearity statistic of Eq.~\eqref{eq:non_additivity}. We observe that for $\alpha \leq 1$, all mass bins are approximately additive and linear. But outside of this, the lower mass halos become non-linear, reaching a maximum of $\epsilon\sim0.3$ for the lowest mass bins and largest radial factors. The galaxy clusters remain relatively linear, whereas all other mass bins exhibit a similar trend toward nonlinearity as the radius factor increases. This is likely due to sensitivities to modifications to the two-halo term when applying replacements so far outside the halo. The key differences in the $C_\ell$ and $P(k)$ statistics are that the weak lensing statistic integrates the effects of baryons over various redshifts, weighted by a lensing kernel. While we observe a comparable structure, the baryonic imprints are fundamentally different.

Fig.~\ref{fig:C_ell_cumulative_summary_z1} condenses this scale-dependent information into the multipole-averaged response fraction $\langle F_{C_\ell} \rangle_\ell$, computed analogously to $\langle F_P \rangle_k$ with Eq.~\eqref{eq:weighted_mean_response}. We average over all multipoles where $|C_\ell^{\rm H} - C_\ell^{\rm DMO}|/C_\ell^{\rm DMO} > 0.01$ and weigh each bin by the relative importance computed by the DMO to hydrodynamical differences and the scatter across the $2000$ realizations. The bottom right corner shows that lowering the mass threshold from $M_{\rm min} =10^{13} \,h^{-1}{\rm M}_\odot$ to $M_{\rm min } =10^{12}\,h^{-1}{\rm M}_\odot$ at fixed $\alpha_{\rm max} = 1$ increases $\langle F_{C_\ell} \rangle_\ell$ from $\sim 0.51$ to $\sim 0.77$—a dramatic gain indicating that low-mass halos cannot be ignored in their importance to the angular power spectrum. Similarly, extending the radius from $\alpha=1$ to $\alpha=3$ yields only modest improvement.

Replacing halos with $M \geq 10^{12}\,h^{-1}{\rm M}_\odot$ out to $\alpha=1$ already captures $\langle F_{C_\ell} \rangle_\ell \sim 0.84$—approximately 84\% of the baryonic effect on the angular power spectrum at $z_s \sim 1$. To achieve $\langle F_{C_\ell} \rangle_\ell \gtrsim 0.92$, one must extend to at least $\alpha=5$ for halos above $M_{\rm min} \geq10^{12}\,h^{-1}{\rm M}_\odot$. This is a very similar requirement to that for the matter power spectrum, which required $\alpha=5$ to approach 91\% (Fig.~\ref{fig:response}).

The upper left corner of the matrix in Fig.~\ref{fig:C_ell_cumulative_summary_z1} shows the discrete models in a similar way to Fig.~\ref{fig:response}. One key difference between the matter power spectrum and the angular power spectrum is that the latter is less sensitive to galaxy clusters and more sensitive to galaxy groups and intermediate-mass halos. We also see that the baryons in the exteriors of the halos have a larger influence on $C_\ell$ than on $P(k)$, which can naturally be explained by the fact that in lensing, the halo cores are diluted by projections along the line of sight. 

We have also computed the redshift evolution for the baryonic effects on $C_\ell$, but find a weak redshift dependence. Unlike the matter power spectrum, the convergence power spectrum response does not show a clear signature of low-mass halos “turning off” their feedback at high redshift. This is again due to projection effects: even though individual low-mass halos at $z \sim 2$ may exhibit weak feedback effects on the power spectrum (as seen in Fig.~\ref{fig:power_spectrum_z_evolution}), the lensing signal for high-$z_s$ sources is dominated by galaxy groups where feedback is more pronounced
%Fig.~\ref{fig:Cl_z_evolution} extends this analysis across source redshifts from $z_s = 0.03$ to $z_s = 2.57$. The layout mirrors Fig.~\ref{fig:power_spectrum_z_evolution}, with each panel showing $\langle F_{C_\ell} \rangle_\ell$ as a function of source redshift for $\alpha_{\rm max}$ and varying $M_{\min}$ in the top row and for discrete bins, $(M_{a}, M_{a+1};\alpha_{i}, \alpha_{i+1})$, in the bottom row. The convergence power spectrum response shows much weaker redshift dependence than the matter power spectrum. Unlike the matter power spectrum, the convergence power spectrum response does not show a clear signature of low-mass halos “turning off” their feedback at high redshift. This is again due to projection effects: even though individual low-mass halos at $z \sim 2$ may exhibit weak feedback effects on the power spectrum (as seen in Fig.~\ref{fig:power_spectrum_z_evolution}), the lensing signal for high-$z_s$ sources is dominated by galaxy groups where feedback is more pronounced.

These results demonstrate that precisely modeling baryonic effects on the weak lensing convergence power spectrum for Stage-III and Stage-IV surveys requires capturing halos down to at least $M \geq 10^{12}\,h^{-1}{\rm M}_\odot$, and extending halo modifications to at least $\alpha_{\rm max}=3$, and preferably capturing two-halo and nonlinear effects from $\alpha_{\rm max}=5$.

\subsection{Beyond Gaussian statistics}
\begin{figure*}[!t]
\centering
\includegraphics[width=\linewidth]{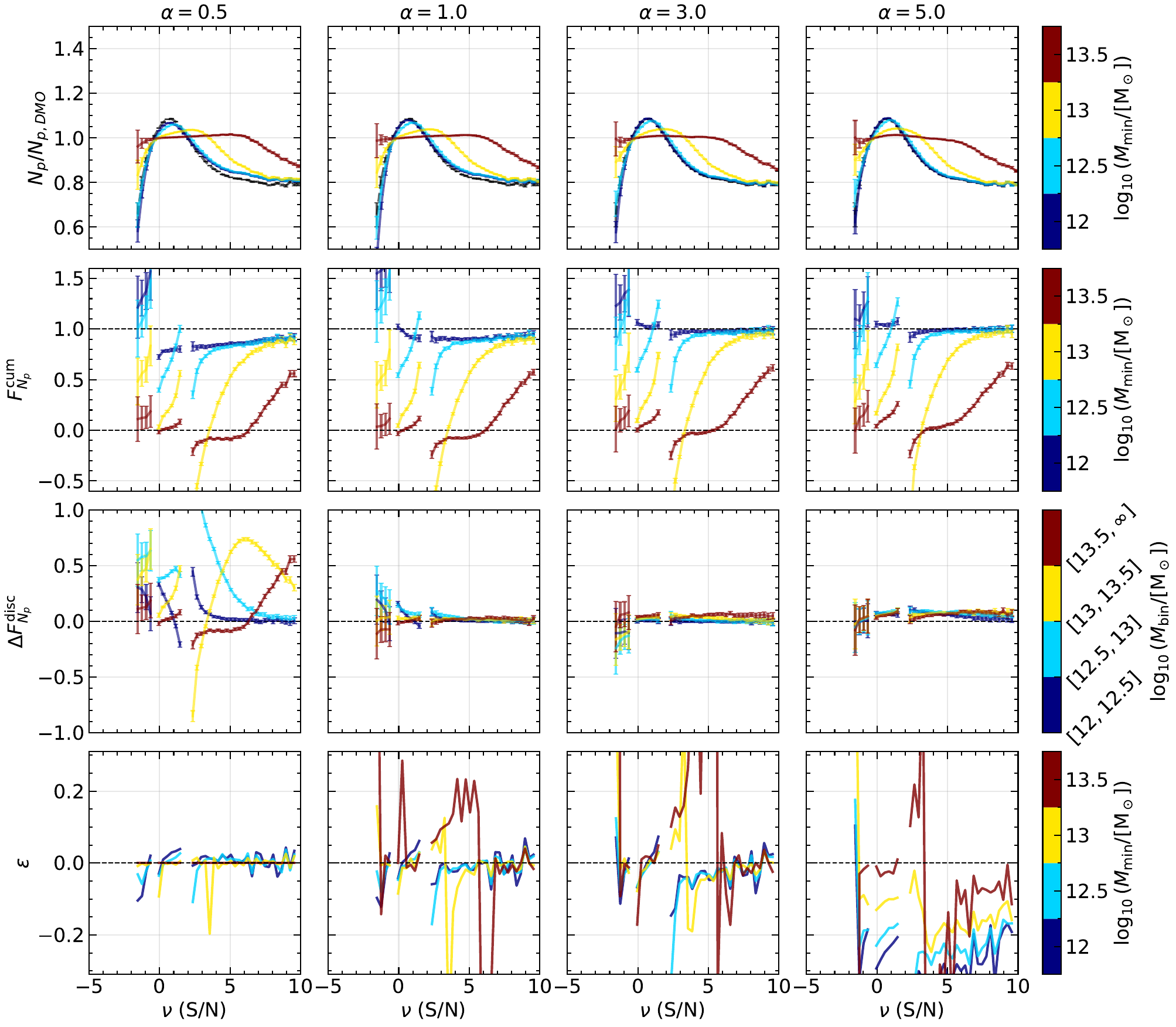}
\caption{Response of convergence peak counts to baryonic effects as a function of signal-to-noise ratio $\nu$ at source redshift $z_s \approx 1.0$ following the same layout and color scheme as Fig.~\ref{fig:suppression_and_fraction} and Fig.~\ref{fig:C_ell_cumulative_summary_z1}. Note that in rows 2-4, the gaps indicate regions where the true hydro-to-dmo difference in peak counts is less than 1\%, so they are excluded from the analysis and presentation.}
\label{fig:peaks_response}
\end{figure*}

We now extend our response analysis to non-Gaussian statistics of the weak-lensing convergence field, which probe the matter distribution in fundamentally different ways from the power spectrum.  For clarity of text, we present detailed results for peak counts in the manuscript's body; analogous analyses for minima, the PDF, and Minkowski functionals are provided in Appendix~\ref{appendix:NGstats}.

For each of the $5.28$ million convergence maps, we compute the statistics of \S~\ref{sec:statistics} on the unsmoothed, noise-free convergence field. We bin peak, and minima counts in 50 linearly spaced $\nu$ bins spanning $-5 \leq \nu \leq 10$. The cumulative and discrete response fractions are defined analogously to the matter and angular power spectrum. As with the $C_\ell$ and $P(k)$, we compute the response statistic for the discrete models with $(M_a, M_{a+1};\alpha_{i}, \alpha_{i+1})$ as well.

Fig.~\ref{fig:peaks_response} shows the cumulative response of peak counts to baryonic effects as a function of $\nu$ at source redshift $z_s \approx 1.0$ paralleling the displays of Fig.~\ref{fig:suppression_and_fraction} and Fig.~\ref{fig:C_ell_cumulative_summary_z1}. As in the previous sections, we again see that baryonic effects on peak counts are primarily sourced by halo cores. The leftmost panel ($\alpha_{\rm max} = 0.5$) shows that replacing only the innermost $0.5R_{200}$ of halos $M\geq10^{12}\,h^{-1}\,{\rm M}_\odot$ achieves $F_{\rm peak} \sim 0.6$–$0.8$ for high-$\nu$ peaks ($\nu \gtrsim 4$), capturing 60–80\% of the total baryonic effect. High-$\nu$ convergence peaks are thought to be sourced by individual massive halos whose lensing signal is determined primarily by the projected density within $\sim R_{200}$ \citep{Yang-13, liu16} and baryonic processes that contract or expand the core density profile, such as AGN feedback evacuating gas from cluster centers or stellar feedback redistributing baryons in groups.

Baryonic effects on peak statistics also appear to saturate at smaller radii than the power spectrum. By $\alpha_{\rm max} = 3$ (third panel), $F_{\rm peak} \gtrsim 0.95$ across nearly all $\nu$, indicating that baryonic modifications beyond $3R_{200}$ contribute negligibly to peak counts. Extending to $\alpha_{\rm max} = 5$ (the fourth panel) yields a marginal improvement. This contrasts with the power spectrum (Fig.~\ref{fig:response} and Fig.~\ref{fig:C_ell_scale_dependent_response_z1}), which required $\alpha_{\rm max} = 5$ to approach $\sim90\%$ fidelity. The difference arises because peak counts probe the local density contrast around individual halos. In contrast, the power spectrum integrates over all Fourier modes, including those sensitive to large-scale correlations between halos and their extended outskirts.

In the discrete view of peaks in the third row, the innermost core region $0 \leq r \leq 0.5R_{200}$ (left panel) exhibits clear mass structure in $\nu$-space: low-mass halos ($10^{12}\,h^{-1}{\rm M}_\odot\leq M < 10^{13}\,h^{-1}{\rm M}_\odot$, purple/blue curves) dominate the response at $\nu \sim 2.5$–$4$, groups ($10^{13}\,h^{-1}{\rm M}_\odot\leq M <10^{13.5}\,h^{-1}{\rm M}_\odot$, yellow) control $\nu \sim 5$–$8$, and clusters ($M \geq 10^{13.5}\,h^{-1}{\rm M}_\odot$, red) dominate the highest peaks $\nu \gtrsim 8$. The middle and right panels reveal a qualitatively different behavior for halo outskirts ($\alpha \geq1$): the discrete response $\Delta F_{\rm peak}^{\rm bin}$ is approximately constant in $\nu$ at $\sim 0.05$ for all radial shells beyond $\alpha=1$. This indicates that gas redistribution in halo outskirts produces a broad, nearly uniform effect on peak counts across all $\nu$, rather than selectively influencing specific peak heights. The most significant effect comes from cluster outskirts ($M \geq 10^{13.5}\,h^{-1}{\rm M}_\odot$, red curves in second from the right panel), where AGN-driven gas redistribution at $1\leq \alpha<3$ contributes $\Delta F_{\rm peak}^{\rm bin} \sim 0.1$ uniformly.

\begin{figure*}[!t]
\centering
\includegraphics[width=\linewidth]{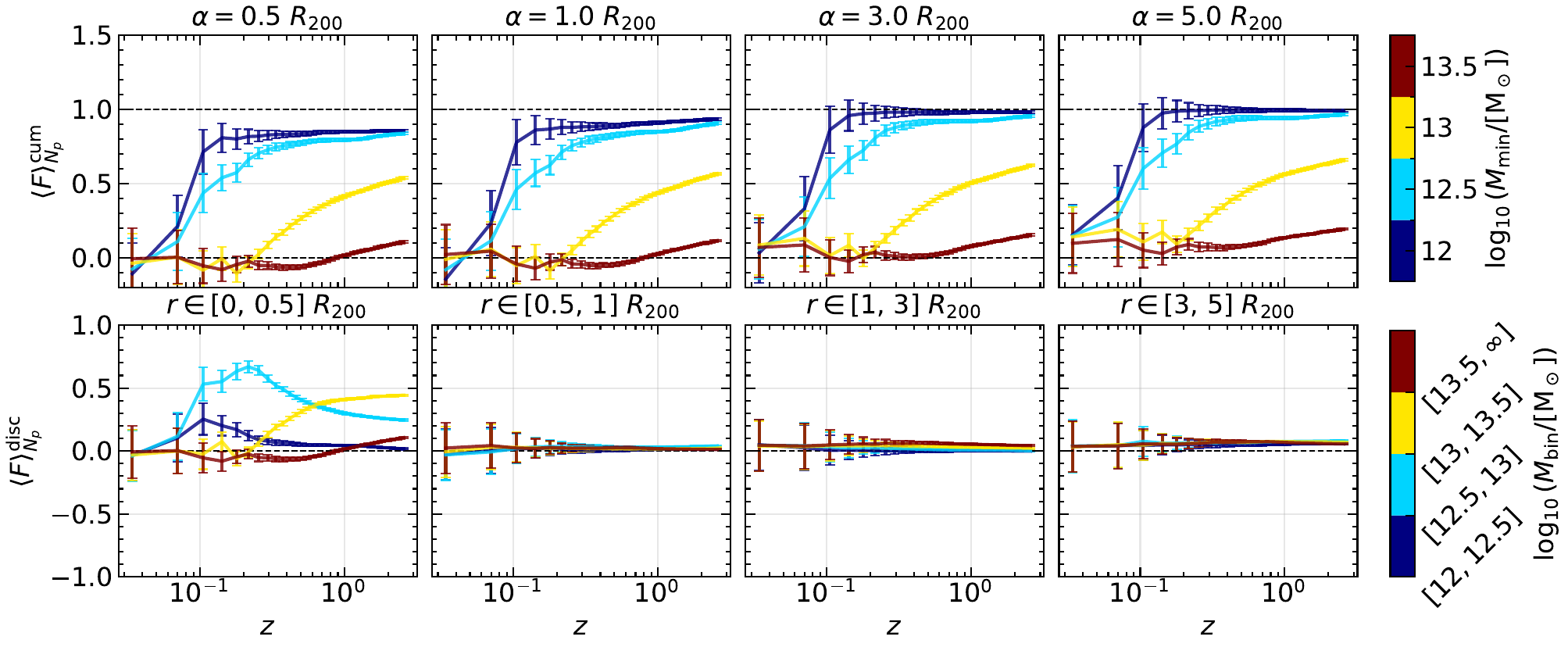}
\caption{Redshift evolution of the weighted mean peak response of Fig.~\ref{fig:peaks_response} averaged over signal-to-noise bins following the same format as Fig.~\ref{fig:power_spectrum_z_evolution}.}
\label{fig:peaks_redshift_evolution}
\end{figure*}

Fig.~\ref{fig:peaks_redshift_evolution}  shows the redshift evolution of the baryonic impact on peaks analogous to Fig.~\ref{fig:power_spectrum_z_evolution}. The left panels ($\alpha_{\rm max} =0.5$) show that at low source redshifts ($z_s \lesssim 0.5$), low-mass halos dominate the peak response, with $\langle \Delta F_{\rm peak}^{\rm bin} \rangle_\nu \sim 0.4$ for $10^{12.5}$–$10^{13}\,h^{-1}{\rm M}_\odot$. Conversely, at high source redshifts ($z_s \gtrsim 1.5$), only more massive halos contribute, with $\langle \Delta F_{\rm peak}^{\rm bin} \rangle_\nu \sim 0.2$ for $M \geq 10^{13.5}\,h^{-1}{\rm M}_\odot$ and $\langle \Delta F_{\rm peak}^{\rm bin} \rangle_\nu \sim 0.35$ for the two mass bins $10^{12.5}$–$10^{13}\,h^{-1}{\rm M}_\odot$, and $10^{13}$–$10^{13.5}\,h^{-1}{\rm M}_\odot$. Outer radial shells ($\alpha > 0.5$) exhibit much weaker redshift evolution, with $\langle \Delta F_{\rm peak}^{\rm bin} \rangle_\nu$ remaining roughly constant at $\sim 0.05$–$0.15$ across all $z_s$. This insensitivity to source redshift indicates that gas at $r \sim R_{200}$–$5R_{200}$ contributes a smooth convergence signal that uniformly boosts the peak counts. In all peak count analyses, halo cores dominate the response to baryonic effects, but do not fully capture them. For that, we would require correct modifications out to $\alpha_{\rm max}=3$. 

% One can use the above information to cross-validate BCMs. The distinct $\nu$-$M$ structure (third row of Fig.~\ref{fig:peaks_response}, left panel) means that BCMs can be tested by checking whether they reproduce the correct peak counts at different $\nu$ thresholds. A model that fits high-$\nu$ peaks ($\nu \gtrsim 8$, cluster-dominated) but fails at $\nu \sim 3$–$5$ (group-dominated) reveals inconsistent treatment of feedback across mass scales.

\begin{figure*}[!t]
\centering
\includegraphics[width=\linewidth]{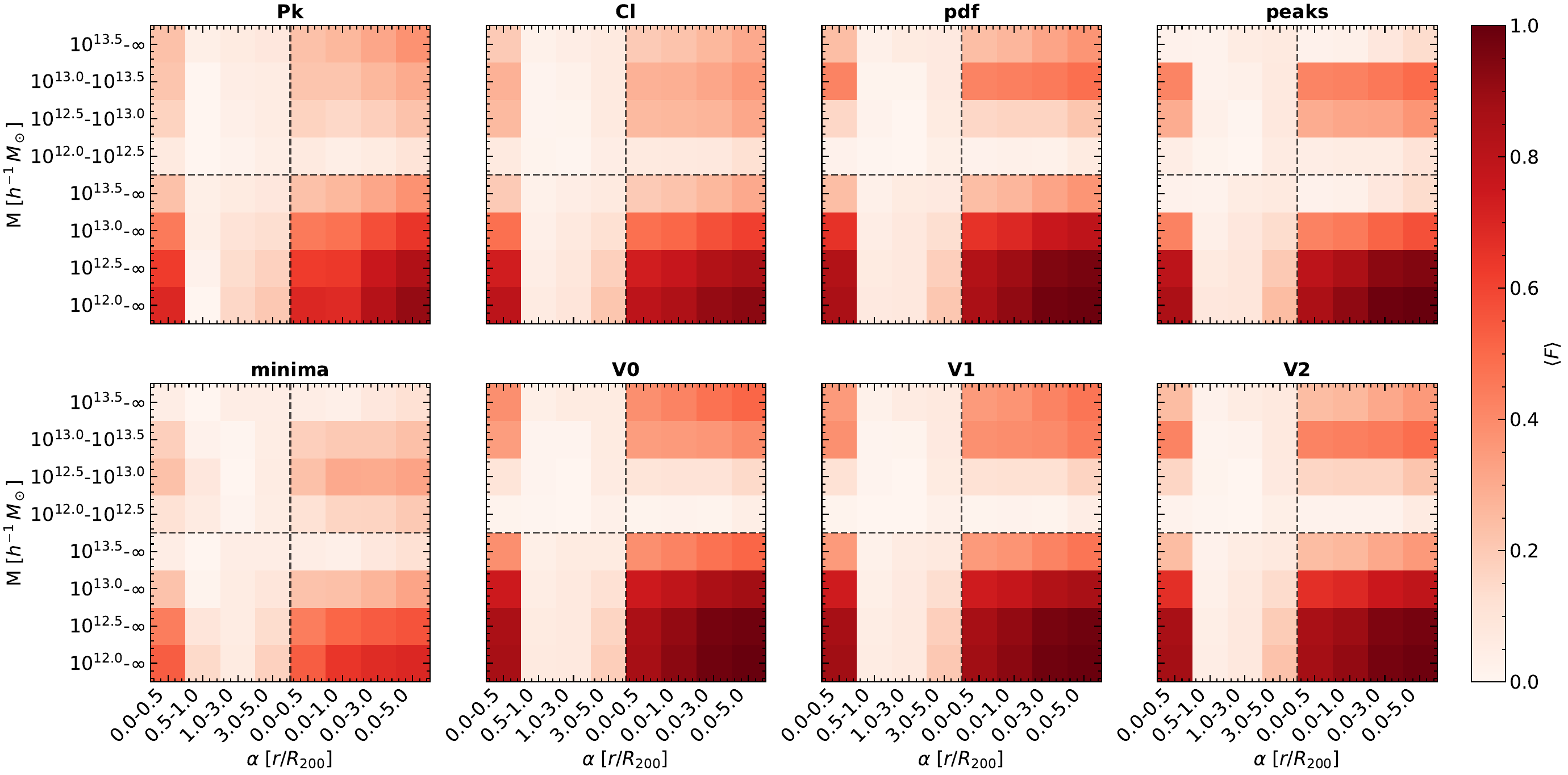}
\caption{Mean response $\langle F_S(M_{\min}, \alpha) \rangle$ across various statistics and for all models at source redshift $z_s \sim 1.0$. As in Fig.~\ref{fig:models}. Each panel corresponds to a different statistic while color encodes the weighted mean response averaged over the relevant binning variable (e.g., $\nu$ for peaks/minima, $\ell$ for $C_\ell$) with weights following Eq.~\eqref{eq:weighted_mean_response}. Most statistics fail to fully reach $\langle F\rangle=1$ even with the most aggressive Replacement, showing percent or greater deviations. This indicates that halo-based Replacement cannot capture all baryonic effects. The centers of halos are the most important for all statistics, however, the mass dependencies change, and the structure of each statistics fingerprint changes}
\label{fig:all_Statistics_heatmap}
\end{figure*}

In Appendix~\ref{appendix:NGstats}, we present response analyses analogous to peaks for minimum counts, the one-point PDF, and Minkowski functionals. Here, we synthesize these results in a comparative heatmap that summarizes the main findings of this section. Fig.~\ref{fig:all_Statistics_heatmap} compares the weighted mean response $\langle F_S(M_{\min}, \alpha_{\rm max}) \rangle$ across eight statistics at source redshift $z_s \approx 1.0$ following the structure of Fig.~\ref{fig:response} and Fig.~\ref{fig:C_ell_cumulative_summary_z1}. We see that even the most aggressive halo Replacement fails to fully reproduce baryonic effects in most statistics. The bottom-right cells of each heatmap—corresponding to replacing all halos with $M_{\rm min} > 10^{12}\,h^{-1}{\rm M}_\odot$ out to $\alpha_{\rm max} = 5$—achieve $\langle F_S \rangle =1.0$ only for peak counts.

Fig.~\ref{fig:all_Statistics_heatmap}  shows that different statistics exhibit different saturation scales in the $M-R$ space. Peak counts achieve $\langle F_{\rm peak} \rangle =1$ by $\alpha_{\rm max} = 3$ for $M_{\min} = 10^{12}\,h^{-1}{\rm M}_\odot$, whereas the convergence power spectrum requires extension to $\alpha_{\rm max} = 5$ to reach $\langle F_{C_\ell} \rangle \sim 0.95$. Minimum counts fall between these extremes, saturating around $\alpha_{\rm max} = 3$–$5$ with $\langle F_{\rm min} \rangle \sim 0.69$ for $M>10^{12}\,h^{-1}{\rm M}_\odot$. The power spectrum exhibits the most gradual saturation, with visible color gradients persisting across the range of $\alpha_{\rm max}$ values.

Statistics that fail to reach $F_S = 1$ even with perfect field-level replacements suggest physical mechanisms beyond our modeling framework. Possibilities include: (1) baryonic effects in halos below our mass threshold ($M < 10^{12}\,h^{-1}{\rm M}_\odot$). These mass ranges were excluded from this work primarily because low-mass halos are not included in standard baryon-correction models. This is typically the case because the physically motivated BCM parameters can be fit to X-ray and SZ observations, but this limits them to higher-mass halos. (2) The intergalactic medium (IGM), such as gas in filaments or galactic outflows extending beyond 5 times the virial radii of halos. The fact that peak counts do reach $F_{\rm peak} \approx 1$ suggests they are uniquely insensitive to these diffuse components, consistent with their status as local, halo-centric statistics \citep{Yang-13, liu16}.

In Fig.~\ref{fig:all_Statistics_heatmap}, we see that baryonic effects in cores source the majority of the impact on most statistics. However, the importance and mass scales differ across statistics. Minimum counts show enhanced sensitivity to low-mass halo cores compared to peaks, while Minkowski functionals exhibit the most significant sensitivity to the massive galaxy clusters.

At the same time, the systematic rise of $\langle F_S\rangle$ along the cumulative-radius direction shows that the cores are not sufficient by themselves. Moving from discrete-core to cumulative-in-radius configurations increases the response for two-point statistics and for the lower-threshold non-Gaussian observables, indicating that the baryonic mass redistributed into halo outskirts also contributes non-negligibly to the total signal. The resulting two-dimensional patterns therefore encode both aspects: a universal reliance on the inner regions of massive halos, together with a statistically dependent sensitivity to the radial extent and mass range over which baryonic redistribution must be modeled accurately. These ``fingerprints" provide a compact, visual summary of where in halo mass–radius space each statistic derives its baryonic information, and thus serve as a practical target for testing and calibrating baryonic correction models that aim to be self-consistent across multiple observables. 

While Fig.~\ref{fig:all_Statistics_heatmap} summarizes the responses in our weighted formalism following Eq.~\ref{eq:weighted_mean_response}, we show the raw, scale dependent percent deviations for each statistic in Appendix~\ref{app:statistical_deviations_from_hydro}, which connects these fingerprints to the more familiar percent level comparisons.

\section{Baryon correction model consistency tests}
\label{sec:BCM}
\begin{figure*}[!t]
    \centering
    \includegraphics[width=\linewidth]{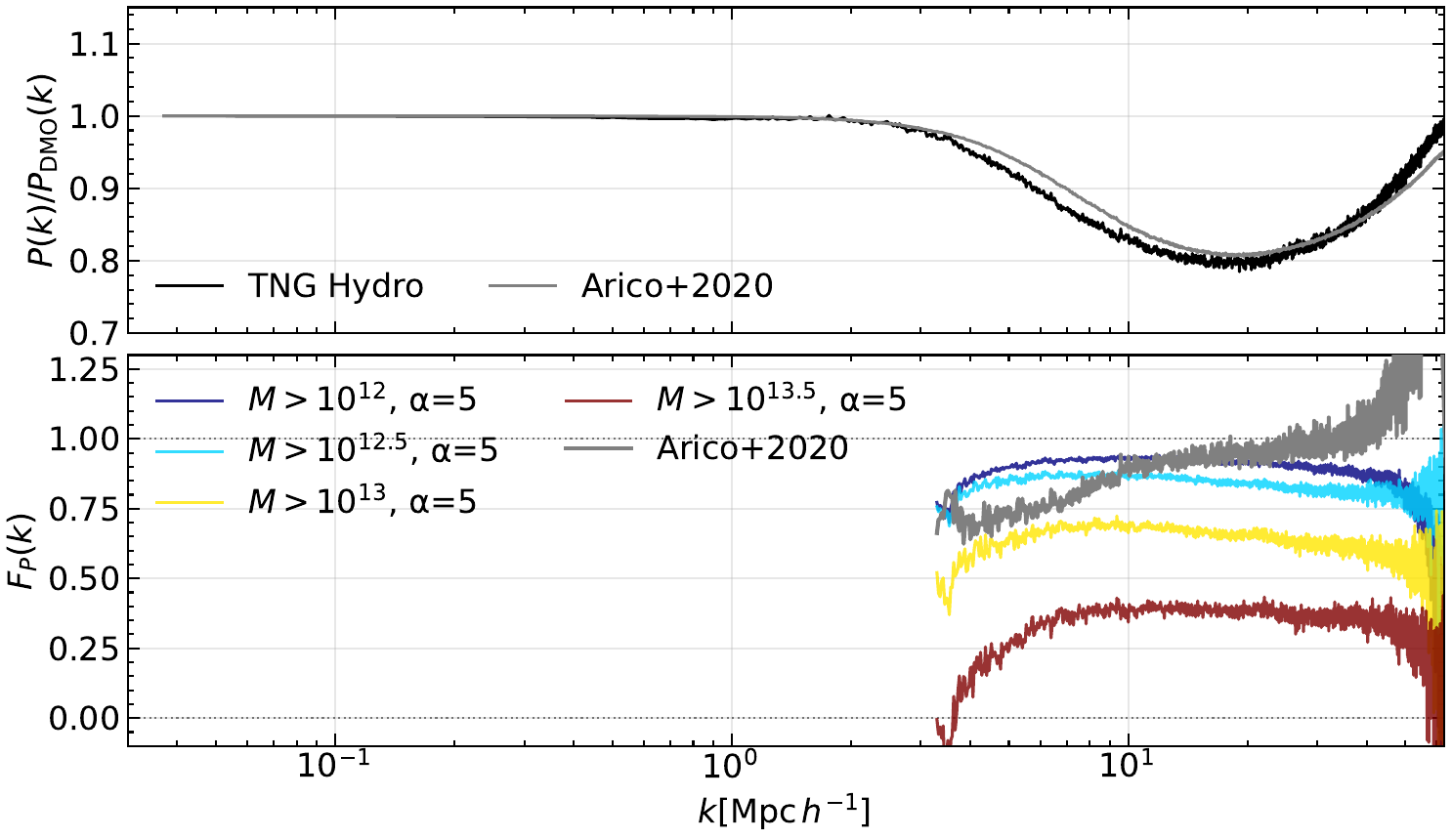}
    \caption{Response fraction comparison between the Arico+2020 BCM and cumulative \textit{Replace} models for the matter power spectrum at $z\sim0.96$. \emph{Top:} The \citet{Arico2020} model (gray) successfully reproduces the target hydrodynamical suppression (black) at percent-level accuracy by design. \emph{Bottom:} The BCM's response fraction $F_P^{\rm BCM}(k)$ (gray) resembles cumulative \textit{Replace} models, sitting closest to \textit{Replace} $M_{\rm min}\geq10^{13}\,h^{-1}\,{\rm M}_\odot$ at intermediate scales ($k\sim3$--10 $h\,{\rm Mpc}^{-1}$) and approaching $M_{\rm min}\geq10^{12}$ at smaller scales ($k\gtrsim10$).}
    \label{fig:F_BCM}
\end{figure*}

The previous sections have provided a detailed overview of how baryons affect various statistics, from the power spectrum of the density and the weak-lensing convergence field to non-Gaussian statistics of the convergence field. We now briefly consider how baryonic correction models (BCMs), which apply spherical particle displacements, may influence statistics and can capture the underlying physical baryonic effects.

BCMs have been shown to reproduce summary statistics such as the matter power spectrum at the few-percent level \citep{Arico2020}, and recent work has demonstrated simultaneous fits to multiple two-point statistics \citep{Arico+2021, Grandis-24}. However, \citet{leeComparingWeakLensing2022} found that the highest significance peaks $\nu\gtrsim4$ were not suppressed enough compared to the true hydrodynamical simulations when using the \citet{Arico2020} BCM fit to the matter power spectrum. Here, we use the response fingerprints from previous sections to reason why that could be the case.

Fig.~\ref{fig:F_BCM} shows the matter power spectrum suppression from IllustrisTNG at $z\sim0.96$ alongside the \citet{Arico2020} best-fit model (applied to all TNG-DM halos with $M\geq2\times10^{12}\,h^{-1}\,{\rm M}_\odot$ following the original implementation in that paper). The top panel confirms percent-level agreement between BCM and target hydro suppression, as expected.

The bottom panel presents the response following the methods in \S~\ref{sec:formalism}, comparing $F_P^{\rm BCM}(k)$ to cumulative \textit{Replace} models. The BCM achieves $F_P \sim 0.8$ across $k\gtrsim 3$, closely tracking the \textit{Replace} $M_{\rm min}=10^{13}\,h^{-1}\,{\rm M}_\odot$, $\alpha_{\rm max}=5$ model at intermediate scales and approaching $M_{\rm min} =10^{12}\,h^{-1}{\rm M}_\odot$, $\alpha_{\rm max}=5$ at smaller scales. This is consistent with the expectation that the BCM modifies halos down to $M\sim2\times10^{12}\,h^{-1}{\rm M}_\odot$, capturing baryonic contributions to the power spectrum from the mass range where our fingerprint analysis (Fig.~\ref{fig:all_Statistics_heatmap}) shows $P(k)$ to be sensitive.

However, matching the $P(k)$ response does not guarantee correct field-level baryon placement within individual halos or matching individual power spectra. The BCM could achieve $F_P^{\rm BCM} \sim 0.8$ by compensating between regions in mass bin-radial shell space that are degenerate for $P(k)$ but distinguishable by other statistics. 

\begin{figure*}[!t]
\centering
\includegraphics[width=\linewidth]{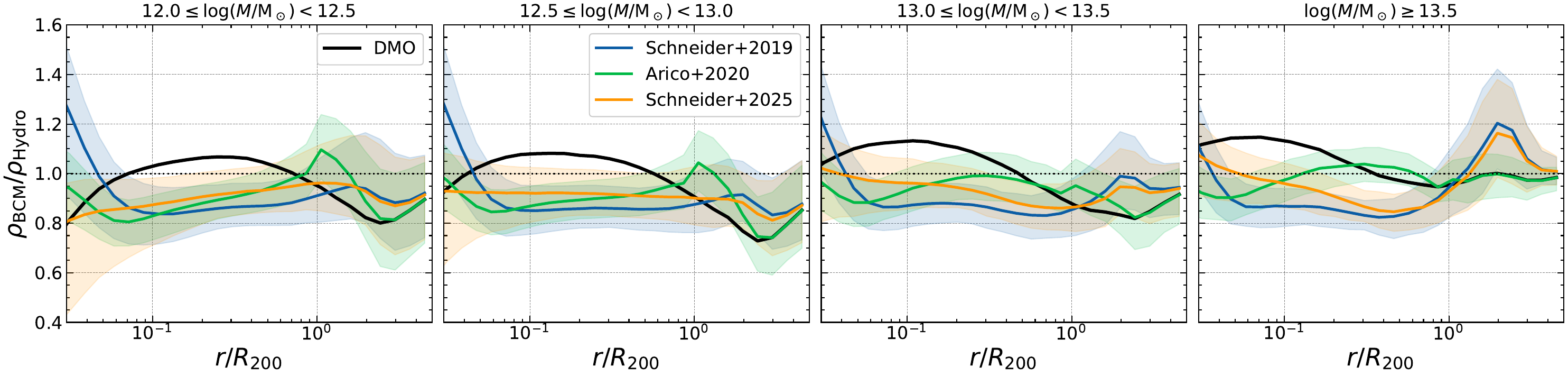}
\caption{Comparison of bijectively matched density profiles from various BCMs, shown as ratios to their hydrodynamical counterpart. Each panel shows a halo mass-bin, with the solid black curves showing the DMO/TNG hydrodynamical ratio. Colored curves show BCM/DMO ratios for: \citep{Schneider19} (blue), \cite{Arico2020} (green), and \cite{Schneider-25} (orange). All BCMs exhibit under-density in cores ($r \lesssim 0.5\,R_{200}$), though the innermost radial bins of the halos appear to be over-concentrated for the Schneider model. The \citet{Arico2020} model shows the closest agreement, but all models exhibit systematic discrepancies that vary with mass and radius.}
\label{fig:bcm_vs_hydro}
\end{figure*}

To test this directly, we compare density profiles from several widely used BCMs \citep{Schneider19, Arico2020, Schneider-25} to bijectively matched TNG profiles. For each model, we apply the standard BCM displacement prescription using the BaryonForge\footnote{\url{https://baryonforge.readthedocs.io/en/latest/}} package \citep{Anbajagane2024bcm} to every bijectively matched halo in the TNG-DM catalog and compute the resulting density profile as a function of radius. Fig.~\ref{fig:bcm_vs_hydro} shows the ratio of BCM-to-Hydro and DMO-to-hydro, averaged in the same mass bins used in the previous sections of this work.

We first observe that all BCMs exhibit underdensity in most halo-core regions ($r \lesssim 0.5\,R_{200}$), with BCM/Hydro ratios systematically below unity. In the innermost bins, the \citet{Schneider19} model shows a significant over-density, but the ensuing under-prediction still outweighs this. We also see in massive halos ($M \gtrsim 10^{13.5} h^{-1} {\rm M}_\odot$) that BCMs systematically overextend the baryon expulsion, showing a noticeable peak at $\sim2R_{200}$. In contrast, the true hydrodynamical profiles converge to the dark matter-only densities by this distance.

\begin{figure*}[!t]
\centering
\includegraphics[width=\linewidth]{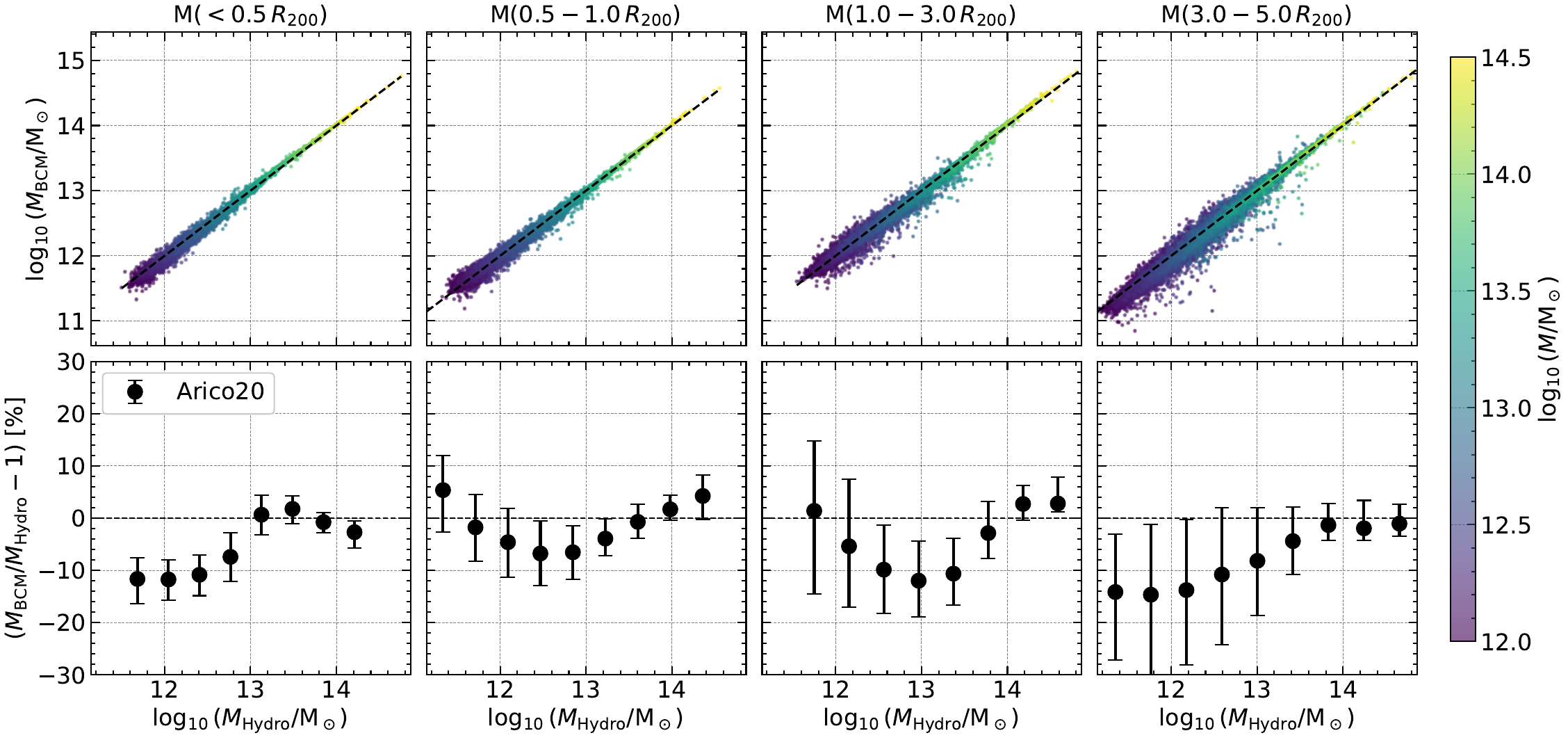}
\caption{Integrated mass comparison between bijectively matched BCM and hydrodynamical halos for various radial shells. \emph{Top:} Enclosed mass $M(r_i<r_o)$ at four representative radii: $0.5R_{200}$, $R_{200}$, $3R_{200}$, and $5R_{200}$. \emph{Bottom:} Fractional residuals $(M_{\rm BCM} - M_{\rm Hydro})/M_{\rm Hydro}$ as a function of halo mass, averaged in mass-bins to highlight the bias. All BCMs under-predict core masses ($r \lesssim R_{200}$) for halos below $\sim10^{13}\,h^{-1}{\rm M}_\odot$ by up to $20\%$, while over-extending the redistribution in massive halos at $r \gtrsim 2R_{200}$.}
\label{fig:shell_mass_comparison}
\end{figure*}

Among the models tested, \citet{Arico2020} shows the closest overall agreement, which is unsurprising given that it is the only model that was fit directly to the IllustrisTNG power spectrum. But this is a nuanced point. While a fit to the power spectrum does not guarantee a match at the profile or field level, it yields a better fit to the density profile than a model not directly tuned to the IllustrisTNG power spectrum. However, even this parameterization exhibits large residuals in both the cores and the outer shells. To highlight this, we show the integrated mass for each profile of bijectively matched BCM and hydro halos, along with their residuals, in Fig.~\ref{fig:shell_mass_comparison}. The top panel shows each halo’s integrated mass, while the bottom plot shows the residual with respect to the 1-to-1 line, clearly showing that for low-mass halos, the model underpredicts core densities by up to 20\%.

In \citet{leeComparingWeakLensing2022}, we found $3\sigma$ discrepancies at the level of peak counts between BCMs and hydrodynamical simulations when BCMs used parameters fit to the matter power spectrum. Taken together with Fig.~\ref{fig:all_Statistics_heatmap}, this is a clear explanation for why. Weak lensing peaks are particularly sensitive to core regions of $10^{12.5}-10^{13.5}\,h^{-1}\,{\rm M}_\odot$ as shown by Fig.~\ref{fig:all_Statistics_heatmap}, and we can clearly see that below $10^{13}\,h^{-1}{\rm M}_\odot$, the BCM model fails to reproduce the cores of these halos, and instead induces a suppression. This systematic core suppression leads to discrepancies for peak counts.

More generally, a BCM calibrated to a single statistic can redistribute baryonic effects among mass bins and radial shells that are degenerate for that statistic but distinguishable by others. For example, fitting to the peaks may induce degeneracies between the $10^{12.5} - 10^{13}\,h^{-1}\,{\rm M}_\odot$ mass bins. On the other hand, the baryonic effects on Minkowski functional $V_1$ is dominated by the higher-mass halos and could break this degeneracy. In  this sense, Fig.~\ref{fig:all_Statistics_heatmap}
 provides a map of which combinations of statistics probe complementary regions of baryonic effects in the underlying field.

Any baryon model that fits the statistics presented here to sub-percent precision must, by construction, violate field-level accuracy at the percent level. Even our \textit{Replace} procedure, which uses the accurate hydrodynamical field-level distribution, falls short by $\sim 10\%$ of the true suppression in $P(k)$ and $C_\ell$ in the most optimistic of replacement models (Fig.~\ref{fig:all_Statistics_heatmap}). This means that a BCM tuned to achieve higher precision must be interpreted as a phenomenological fit to summary statistics, rather than as a faithful reconstruction of baryonic effects. Or said another way, BCMs fit to a given statistic make two mistakes that may cancel each other: they over-fit the effect on halos and ignore the effects outside of their radial influence or below their mass threshold, providing a good fit to the statistic, but an incorrect representation of the field. 
%This distinction becomes critical when combining multiple observables: a BCM tuned to $C_\ell$ will likely predict incorrect peak counts, or minima, if those statistics weight baryonic effects differently across halo masses and radial shells, but not unlikely to fail for a statistic that is sensitive to the same mass bins and radial shells. Or put more clearly, two statistics are generally sensitive to the same field in different ways, and the Fig.~\ref{fig:all_Statistics_heatmap} shows the exact ways in which they differ.

Our \textit{Replace} formalism should then be viewed as an optimistic benchmark for BCMs. It operates at the field level and provides a full non-spherical structure, whereas typical BCMs impose spherical symmetry and neglect contributions from lower mass halos. Even if a hypothetical BCM perfectly reproduced the spherically averaged density profiles of all halos (eliminating the discrepancies in Fig.~\ref{fig:bcm_vs_hydro}), it would still potentially fail to fully recover hydrodynamical statistics, due to both their limited mass coverage and their spherical nature.

% First, our cumulative \textit{Replace} models show that even replacing all matched halos with $M \geq 10^{12}\,h^{-1}{\rm M}_\odot$ out to $\alpha_{\rm max}=5$ captures only $\lesssim 90$\% of the hydrodynamical suppression in $P(k)$ and $C_\ell$ (Fig.~\ref{fig:all_Statistics_heatmap}). BCMs, which operate only on halos, cannot capture extragalactic or low mass components without additional modeling. Second, BCMs impose spherical symmetry when applying radial displacements, whereas true hydrodynamical halos are far more complex, showing, for example, triaxiality and large-scale alignments of outflows with filaments. Our \textit{Replace} fields preserve the complete three-dimensional structure of the hydrodynamical simulation within activated replace mass bins and radial shells. The additional error introduced by spherical averaging is challenging to quantify without running a full BCM through our response pipeline, which we leave for future work. These results, therefore, represent an upper bound. Even a BCM which has been perfectly calibrated to halo profiles in TNG would fall short of the complete \textit{Replace} benchmark, and statistics sensitive to non-spherical structure would see larger discrepancies.

X-ray and SZ observations of group and cluster gas can constrain BCM parameters. However, it is clear that, for virtually all statistics, fitting only to observables from high-mass halos is insufficient. Instead, we must incorporate BCM parameters and models applicable to the lower-mass and IGM regimes.

\section{Conclusions}
\label{sec:conclusion}

In this work, we have introduced a framework for quantifying the origins of baryonic effects on various large-scale-structure statistics. We construct hybrid \textit{Replace} fields that selectively inject hydrodynamical particles to replace dark matter in dark-matter-only simulations across a grid of halo-mass bins and radial shells. By comparing the \textit{Replace} fields to their parent dark matter-only (DMO) simulations, we empirically mapped the statistical response to baryons for the matter and 2D angular weak-lensing convergence power spectra, weak-lensing peak counts, minima, the one-point PDF, and Minkowski functionals. Our key findings are as follows:

\begin{enumerate}
\item Incomplete field-level recovery of baryonic effects sets fundamental limits on statistics. Even replacing all halos above $M\geq10^{12}\,h^{-1}{\rm M}_\odot$ out to $r\leq5R_{200}$ captures only $\sim90\%$ of the hydrodynamical suppression in $P(k)$ and $C_\ell$. The remaining $\sim10\%$ originates from lower-mass halos, and/or material beyond $5R_{200}$, or in the diffuse intergalactic medium, establishing an upper bound on what halo-based correction schemes can achieve.

\item Baryons have a distinct fingerprint on each statistic. The matter power spectrum responds to baryons across halo masses and radial shells, while the effect on peak counts is sharply localized to halo cores ($M\gtrsim10^{12.5}\,h^{-1}{\rm M}_\odot$, $r < R_{200}$). These fingerprints uniquely encode which models accurately predict baryon responses for each statistic.

\item Baryon correction models (BCMs) exhibit systematic spatial biases. When applied to bijectively matched TNG halos, existing BCMs calibrated to $P(k)$ under-predict core masses by $\sim20\%$ for halos below $10^{13}\,h^{-1}{\rm M}_\odot$ while over-extending baryon expulsion in cluster outskirts. These two "mistakes" cancel, yielding a percent-level match for $P(k)$.

\item Because peak counts are sensitive to baryonic contributions at these core regions, BCMs tuned to $P(k)$ systematically fail for peak statistics. This explains discrepancies reported in prior work \citep{leeComparingWeakLensing2022}. 

\item More generally, fitting one statistic may allow BCMs to trade baryonic effects between mass bin and radial shell regions that are degenerate for that observable but distinguishable by other statistics.

\item Matching a statistic at the percent level to hydrodynamical simulations in an unweighted sense can still miss the baryonic signal due to incorrect modelling in regions where baryons matter the most. This implies that BCMs must be tuned specifically to these high-sensitivity scales, rather than merely achieving percent-level residuals with respect to hydrodynamics alone. 
\end{enumerate}

Achieving the percent-level systematic control required by {\it Euclid}, LSST, and {\it Roman} demands either hydrodynamical simulations covering the whole cosmological and subgrid astrophysical parameter space, or models of baryonic effects that can be validated independently against multiple observables with distinct mass-radius sensitivities. Our baryonic fingerprint analysis pinpoints which halo populations must be modeled most accurately for each science case, enabling targeted calibration strategies.

We emphasize that this work exclusively used the IllustrisTNG galaxy formation model. In the future, we hope to apply this framework to alternative hydrodynamical suites (FLAMINGO, SIMBA, EAGLE), which we expect will have different baryonic fingerprints. We can use the methodology presented here to assess dependencies on subgrid models. Similarly, we can apply it to BCMs to rigorously assess their baryon modeling across halo masses and radial shells. The baryonic response formalism provides both a diagnostic tool for existing models and a roadmap for next-generation correction schemes that must satisfy the requirements of next-generation surveys.

\section*{Acknowledgements}

We thank Daniel Angl\'es-Alc\'azar, Greg Bryan, Colin Hill, and Boryana Hadzhiyska for helpful conversations during this work. MEL is supported by NSF grant DGE-2036197. ZH acknowledges financial support from NASA ATP grant 80NSSC24K1093. The Flatiron Institute is supported by the Simons Foundation. The authors used Claude Opus 4.5 to refine sections of code, and Grammarly was used to refine portions of the draft with grammatical edits. The authors take full responsibility for the final content.

\bibliography{biblio}{}
\bibliographystyle{aasjournal}

\appendix
\section{Statistical deviations from hydro}\label{app:statistical_deviations_from_hydro}
\begin{figure}
    \centering
    \includegraphics[width=\linewidth]{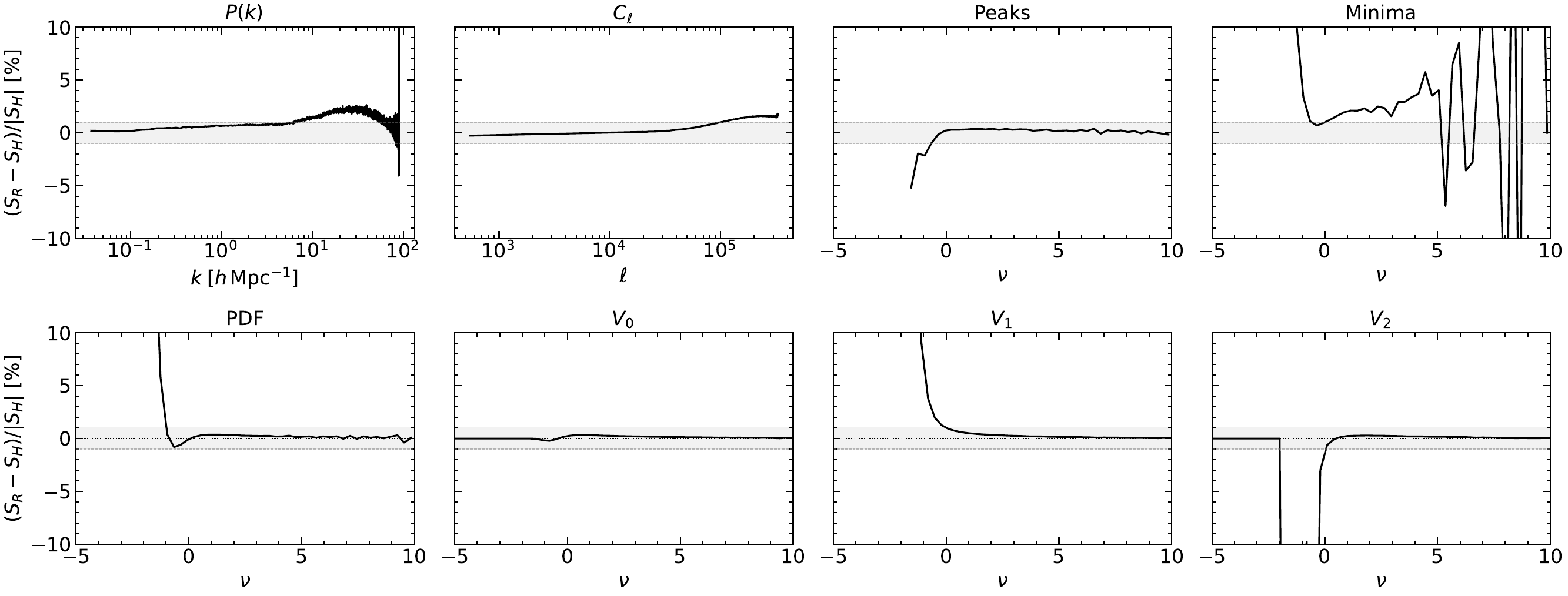}
    \caption{Deviations of the most aggressive \textit{Replace} model ($M\geq10^{12}\,h^{-1}\,{\rm M}_\odot$ and $r<5R_{200}$) from the full hydrodynamical simulation for each of the statistics studied in this work. The grey shaded bands represent $\pm1\%$ deviations. This view of the statistics is complementary to those showing Eq.~\ref{eq:cum_resp_func}, as it removes the implicit weighting by the magnitude of the baryonic effect at each scale.}
    \label{fig:deviation_from_the_truth}
\end{figure}

In \S~\ref{sec:formalism}, we outlined a response formalism that weights a given statistic's response by the importance of baryonic effects (Eq.~\ref{eq:cum_resp_func}). The response represents the deviation of a model from the hydrodynamical simulation weighted by the true baryonic effects. We can rewrite Eq.~\ref{eq:cum_resp_func} then as a relative difference term and a baryonic effect weighting term with,
\begin{equation}
\begin{split}
    F_S &= \dfrac{S_R - S_H}{S_H} B^{-1}\\
    B &= \dfrac{S_D - S_H}{S_H},
\end{split}
\end{equation}
where $S_R$ is the measured statistic from the \textit{Replace} or BCM model, $S_H$ is from the hydrodynamical simulation, and $S_D$ is from the DMO simulation. 

The numbers we quote in the main text are not simply raw fractional residuals given by the first term in the above equation, but deviations of the model's statistics from the true hydrodynamic model, reweighted by the relative strength of baryonic effects at each scale of that statistic ($B^{-1}$). It is, however, common and intuitive to quote the unweighted deviation of a model from the truth at a given scale, e.g., the match of a model to the full hydro for the power spectrum at $k=10\,h\,{\rm Mpc}^{-1}$. To help build intuition for how our weighted response sensitivities relate to the more familiar percent residuals, we show in Fig.~\ref{fig:deviation_from_the_truth} the unweighted fractional deviations for each statistic in our most aggressive replace model where $M\geq10^{12}\,h^{-1}\,{\rm M}_\odot$ and $r<5R_{200}$. 

Fig.~\ref{fig:deviation_from_the_truth} shows that most statistics match the hydrodynamical statistics at the percent level across most scales, with some regions of deviations, such as in the low-$\nu$ PDF. Rather than treating all scales equally as in Fig.~\ref{fig:deviation_from_the_truth}, the response formalism of \S~\ref{sec:formalism} explicitly downweights scales where $B$ is small, namely where baryonic effects are intrinsically weak, and upscales regions where the baryons have the largest impact. Thus, the quoted sensitivities in the main body of this text are the deviations in the most relevant regions for a given statistic. 

As a result, statistics that appear accurate in the raw percent residual with respect to the hydrodynamic simulations can exhibit very different response-weighted sensitivities because baryonic information is concentrated in different regions of scale for each statistic. 

\section{Replacement cuttoff scale effects}
\label{appendix:cutoff}
\begin{figure}
\centering
\includegraphics[width=\linewidth]{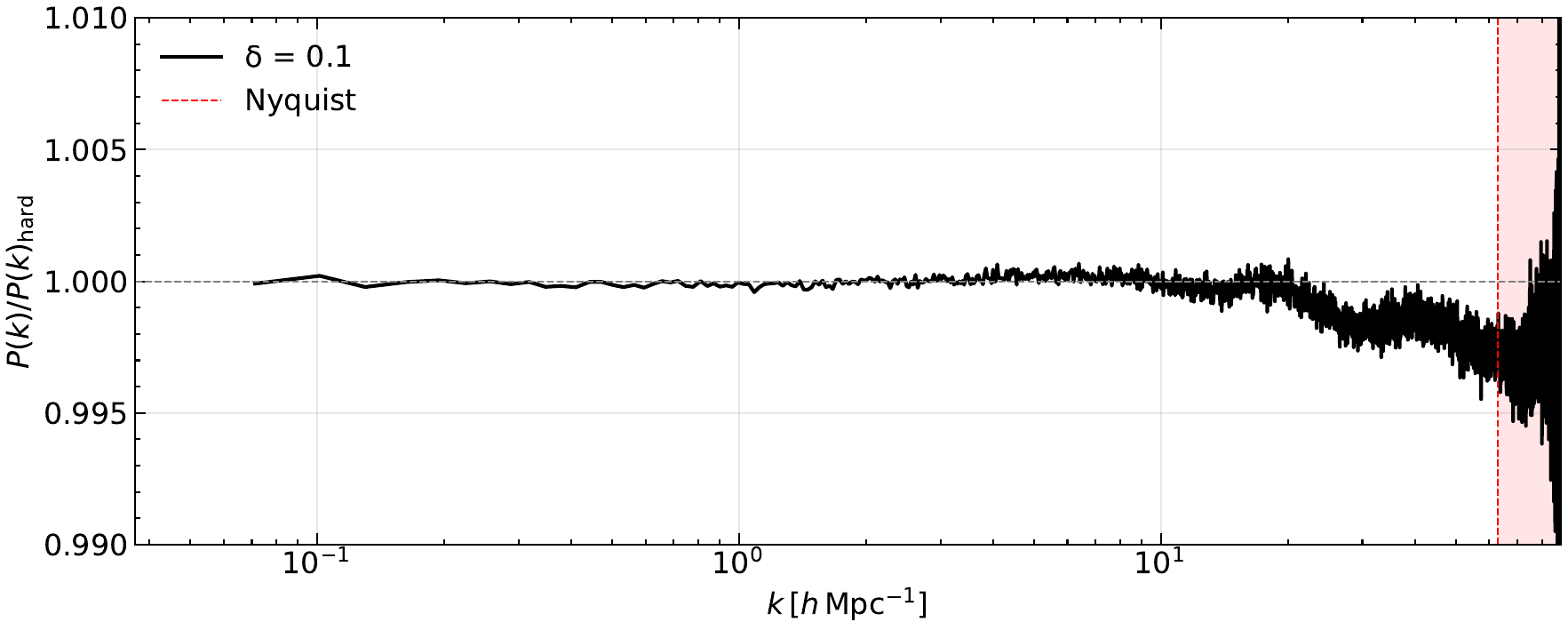}
\caption{Difference in matter power spectrum between hard boundary and smooth cosine ramp Replacement methods for halos with $M>10^{12}\,h^{-1}\,{\rm M}_\odot$ at $\alpha=0.5R_{200}$. The smooth ramping (with $\delta=0.1R_{200}$) yields sub-percent-level differences across all scales used in the analysis, justifying the hard boundary approach adopted throughout this work.}
\label{fig:ramp_test}
\end{figure}
One potential concern is the boundary effects induced by the Replacement methods outlined in \S~\ref{sec:formalism}, as we enforce hard boundaries at the various $\alpha$ levels. To determine whether this will significantly affect our analysis, we perform a simple test. We apply a Replacement for all halos with $M>10^{12}\,h^{-1}\,{\rm M}_\odot$ within $\alpha=0.5R_{200}$. We perform the same Replacement, but instead of a hard boundary, we apply a smooth ramp between dark-matter-only and hydrodynamic particles by weighting particles with a cosine ramp.
\begin{equation}
w(r) = \begin{cases}
0 & r < \alpha R_{200} - \delta \\
\frac{1}{2}\left[1 + \cos\left(\pi\frac{r - \alpha R_{200}}{\delta}\right)\right] & \alpha R_{200} - \delta \leq r \leq \alpha R_{200} + \delta \\
1 & r > \alpha R_{200} + \delta
\end{cases}
\end{equation}
where (w=0) corresponding to the hydrodynamic simulation particles and (w=1) to the dark matter-only particles, with $\delta=0.1R_{200}$ providing the transition width for each halo.

We measure the matter power spectrum for both fields and compare the effects. As shown in Fig.~\ref{fig:ramp_test}, applying smoothing rather than a rigid boundary results in sub-percent changes in the power spectrum across all scales. At the Nyquist frequency, the impacts become somewhat larger, but these scales are removed from all analyses presented in this work. This validates the usage of a Replacement scheme with rigid boundaries as implemented throughout the main text.

\section{Impact of baryons on on other non-Gaussian statistics}
\label{appendix:NGstats}
In \S~\ref{sec:results_WL}, we presented the results for the angular power spectrum and the weak lensing peaks. We limited this presentation for clarity and focus; here, we show the cumulative, discrete, and non-additivity metric for the weak-lensing PDF, minimas, and Minkowski functionals. We follow the same procedure as outlined in \S~\ref{sec:results_WL} for each statistic.

\begin{figure}
    \centering
    \includegraphics[width=\linewidth]{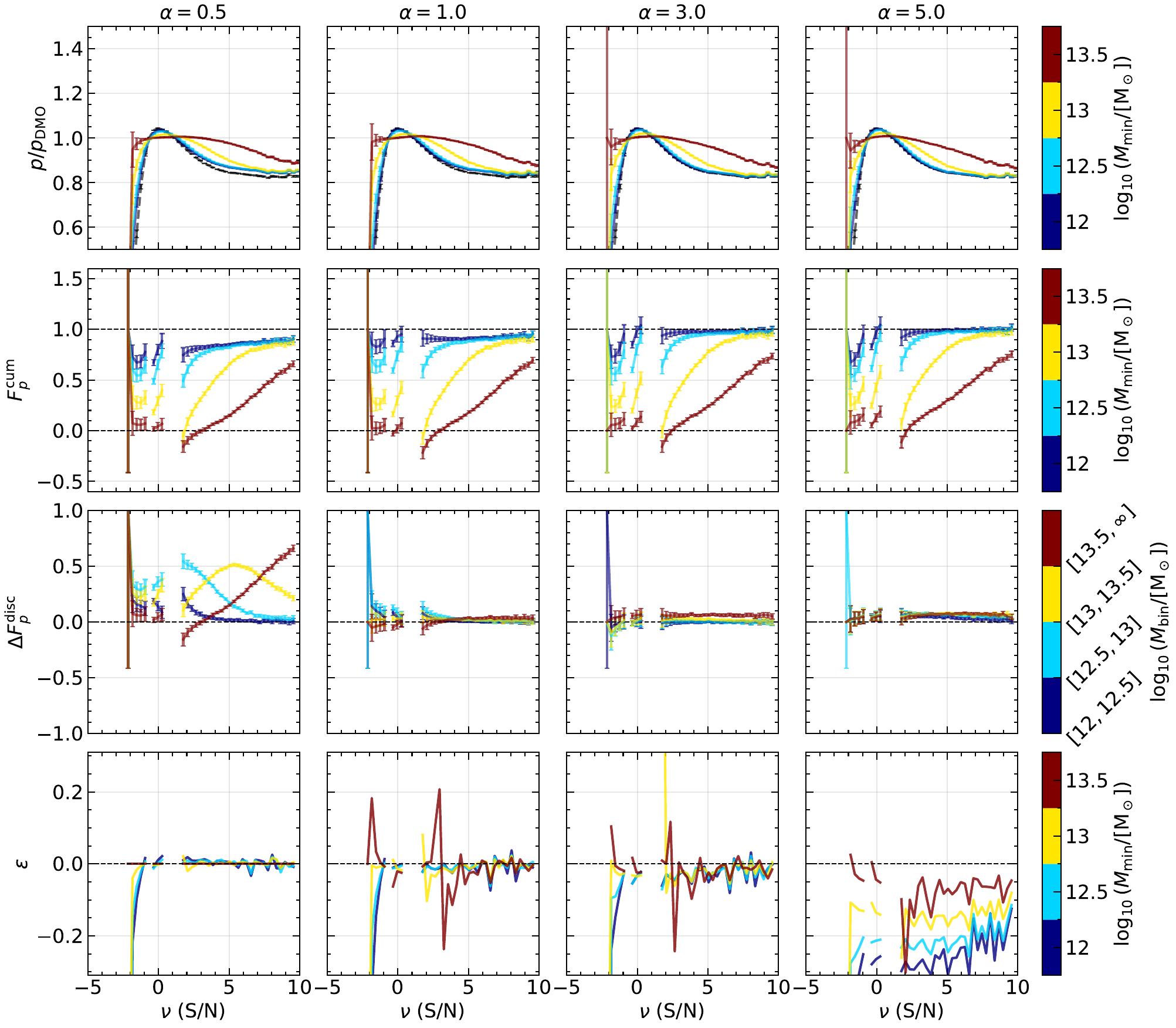}
    \caption{Response of one point PDF to baryonic effects as a function of signal-to-noise ratio $\nu$ at source redshift $z_s \approx 1.0$ following the same layout and color scheme as Fig.~\ref{fig:peaks_response}. Note that in rows 2-4, the gaps indicate regions where the true hydro-to-dmo difference in peak counts is less than 1\%, so they are excluded from the analysis and presentation.}
    \label{fig:pdf_response}
\end{figure}

\begin{figure}
    \centering
    \includegraphics[width=\linewidth]{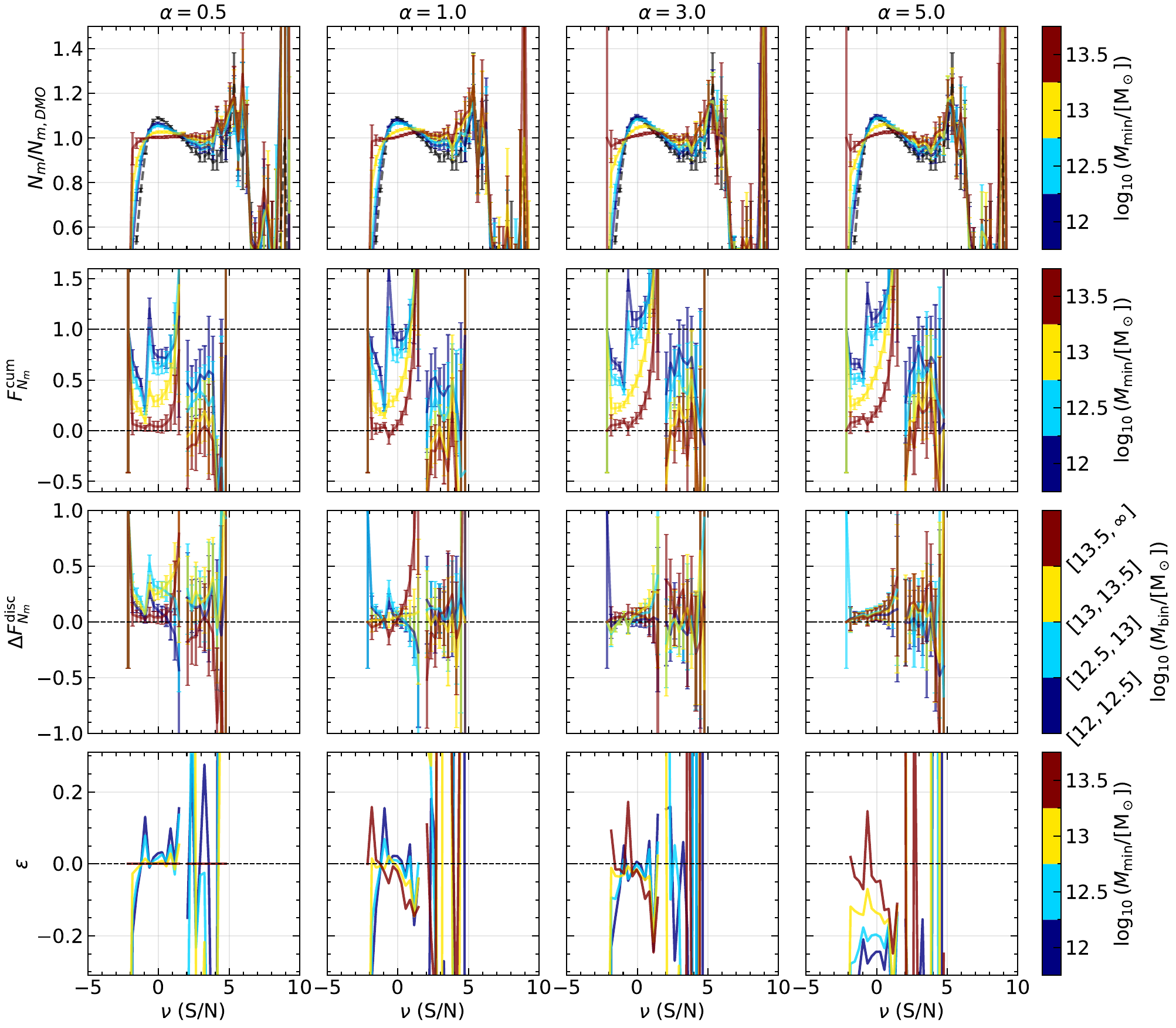}
    \caption{The same as Fig~\ref{fig:pdf_response} but for weak lensing minima}
    \label{fig:minima_response}
\end{figure}

\begin{figure}
    \centering
    \includegraphics[width=\linewidth]{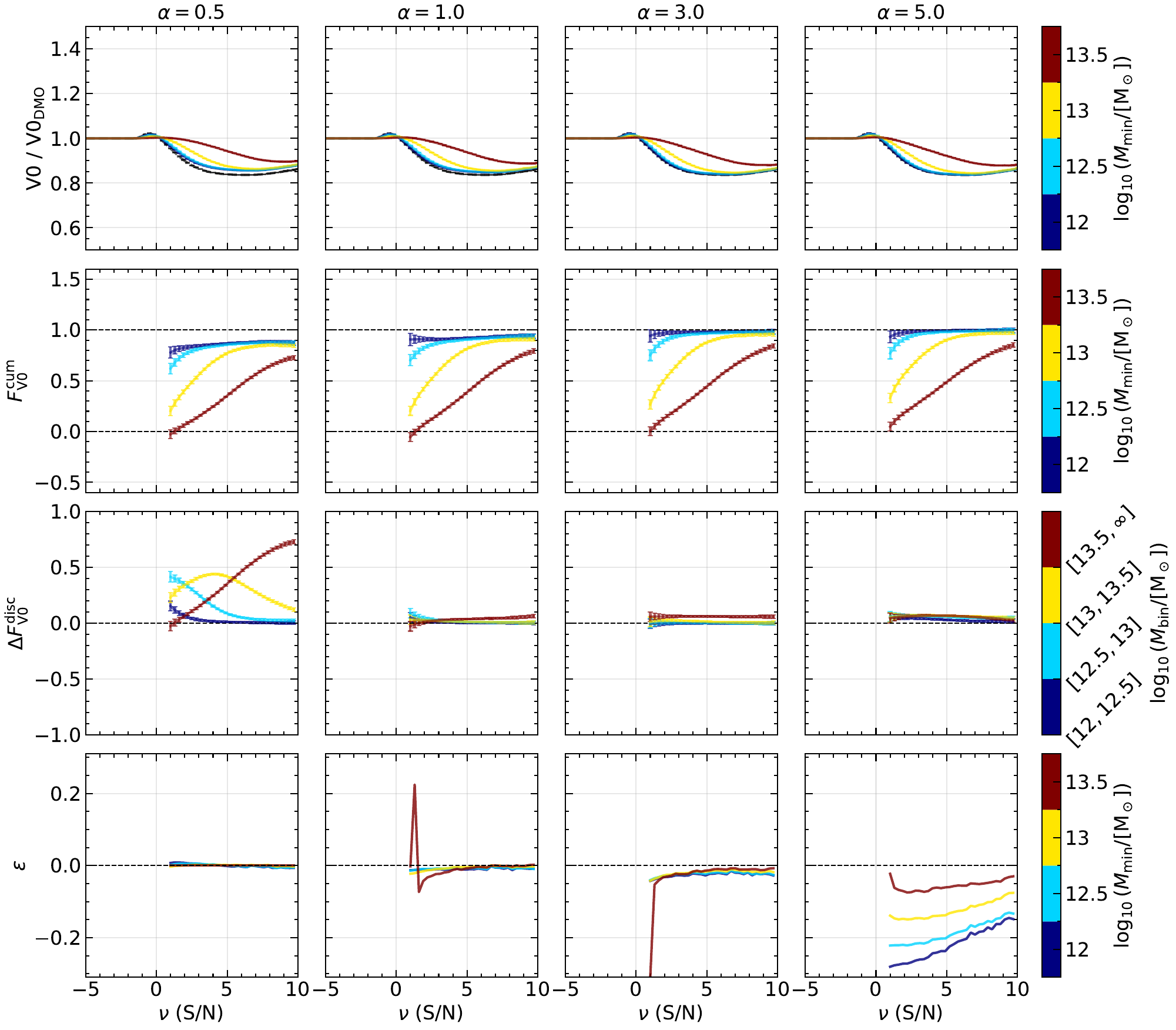}
    \caption{The same as Fig~\ref{fig:pdf_response} but for the first Minkowski functional}
    \label{fig:V0_response}
\end{figure}

\begin{figure}
    \centering
    \includegraphics[width=\linewidth]{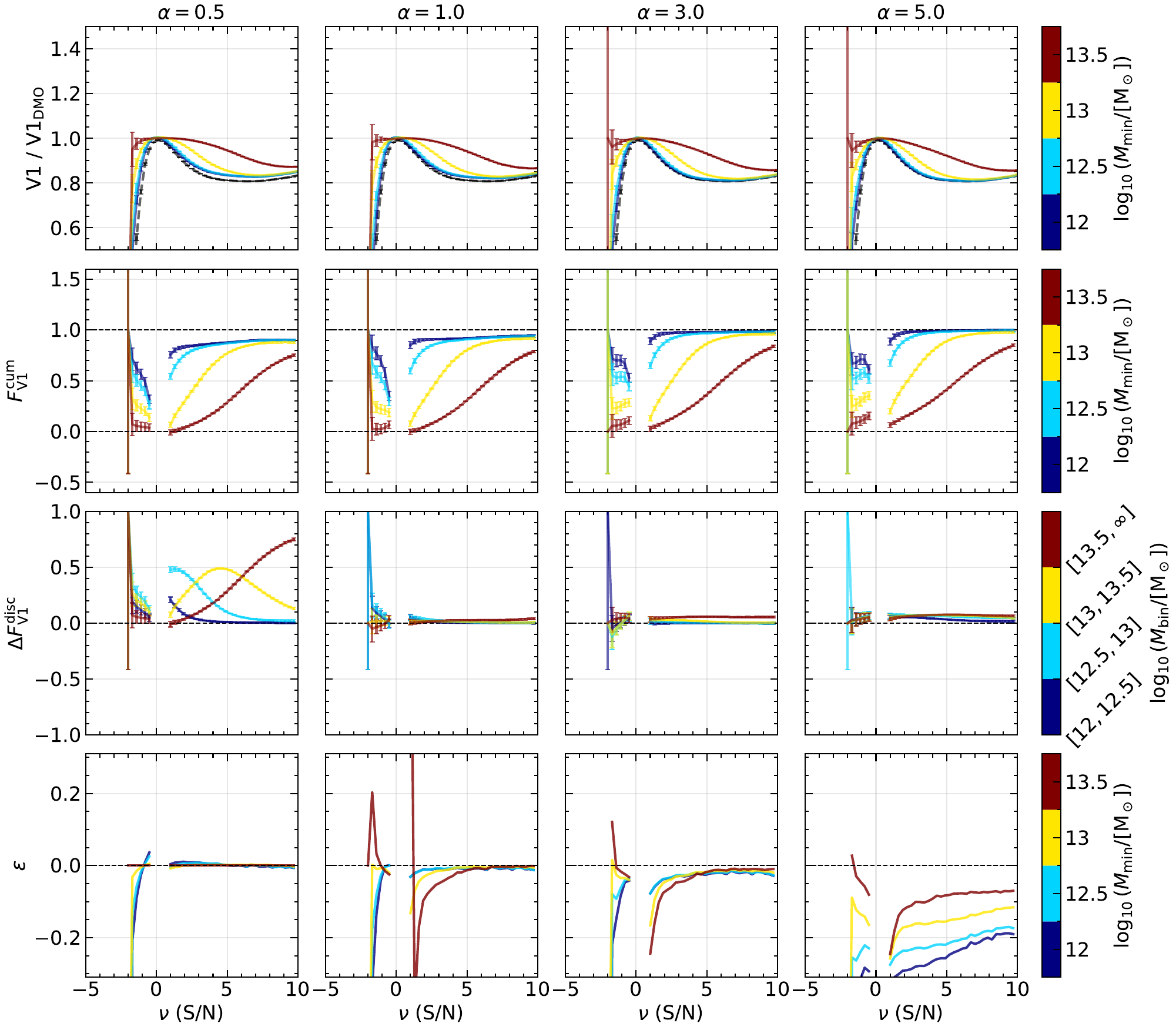}
    \caption{The same as Fig~\ref{fig:pdf_response} but for the second Minkowski functional}
    \label{fig:V1_response}
\end{figure}

\begin{figure}
    \centering
    \includegraphics[width=\linewidth]{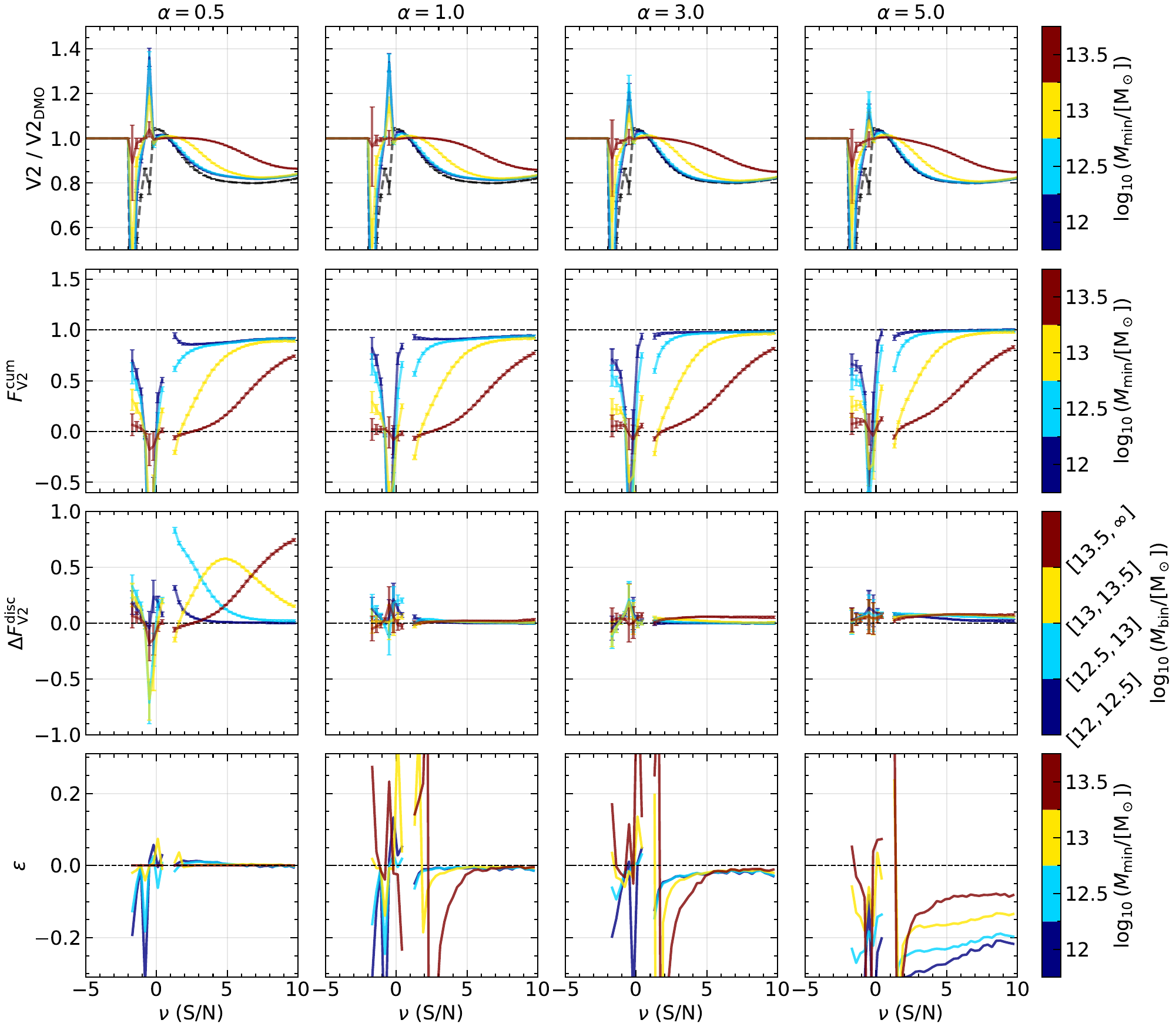}
    \caption{The same as Fig~\ref{fig:pdf_response} but for the third Minkowski functional}
    \label{fig:V2_response}
\end{figure}

In Fig~\ref{fig:pdf_response} we show the response in the same structure and format as Fig.~\ref{fig:peaks_response}. We then follow this structure for Fig.~\ref{fig:minima_response} for minima, and Fig.~\ref{fig:V0_response}, Fig.~\ref{fig:V1_response} and Fig.~\ref{fig:V2_response} for the three Minkowski functionals. In all of the plots, we note that regions of $\nu$ are not shown in the $F$ metric when the true hydro-to-dmo difference is less than $1\%$. 


\end{document}